\pdfoutput=1
\documentclass[10pt,journal]{IEEEtran}

\ifCLASSOPTIONcompsoc
\usepackage[nocompress]{cite}
\else
\usepackage{cite}
\fi
\usepackage{multirow}
\usepackage{xcolor,colortbl}
\definecolor{mygreen}{RGB}{0,255,0}
\definecolor{newgreen}{RGB}{169,208,142}
\definecolor{newred}{RGB}{255,170,161}
\newcommand{\etal}{\textit{et al.}}
\ifCLASSINFOpdf
\usepackage[pdftex]{graphicx}
\else
\fi
\usepackage{amsmath}
\usepackage{array}
\usepackage{siunitx}
\usepackage{dcolumn}
\usepackage{float}
\usepackage{subfig}
\usepackage{comment}
\usepackage{amsmath,mathrsfs}
\usepackage{amsfonts}

\usepackage{hyperref}

\hypersetup{
	colorlinks=true,
	linkcolor=blue,
	filecolor=blue,      
	urlcolor=blue,
	citecolor=mygreen,
}

\newcolumntype{d}[1]{D{.}{.}{#1}}

\begin{document}
	\graphicspath{{./figure/}}
	
	\title{Advances In Video Compression System Using Deep Neural Network: A Review And Case Studies}
	
	\author{Dandan~Ding$^\star$,~\IEEEmembership{Member,~IEEE,}
	    Zhan~Ma$^\star$,~\IEEEmembership{Senior Member,~IEEE,}
		Di~Chen,~\IEEEmembership{Member,~IEEE,}
		Qingshuang~Chen,~\IEEEmembership{Member,~IEEE,}
		Zoe~Liu,
		and~Fengqing~Zhu,~\IEEEmembership{Senior Member,~IEEE}
		\thanks{D.~Ding is with the School of Information Science and Engineering, Hangzhou Normal University, Hangzhou, Zhejiang, China.}
		\thanks{Z.~Ma is with the School of Electronic Science and Engineering, Nanjing University, Nanjing, Jiangsu, China.}
		\thanks{D.~Chen, Q.~Chen, and F.~Zhu are with the School of Electrical and Computer Engineering, Purdue University, West Lafayette, Indiana, USA.}
		\thanks{Z.~Liu is with Visionular Inc, 280 2nd St., Los Altos, CA, USA.}
		\thanks{$^\star$These authors contributed equally.}
    }	

	\maketitle
	
	\begin{abstract}
		Significant advances in video compression system have been made in the past several decades to satisfy the nearly exponential growth of Internet-scale video traffic. From the application perspective, we have identified three major functional blocks including pre-processing, coding, and post-processing, that have been continuously investigated to maximize the end-user quality of experience (QoE) under a limited bit rate budget. Recently, artificial intelligence (AI) powered techniques have shown great potential to further increase the efficiency of the aforementioned functional blocks, both individually and jointly. In this article, we review extensively recent technical advances in video compression system, with an emphasis on deep neural network (DNN)-based approaches; and then present three comprehensive case studies. On pre-processing, we show a switchable texture-based video coding example that leverages DNN-based scene understanding to extract semantic areas for the improvement of subsequent video coder. On coding, we present an end-to-end neural video coding framework that takes advantage of the stacked DNNs to efficiently and compactly code input raw videos via fully data-driven learning. On post-processing, we demonstrate two neural adaptive filters to respectively facilitate the in-loop and post filtering for the enhancement of compressed frames. Finally, a companion website hosting the contents developed in this work can be accessed publicly at \url{https://purdueviper.github.io/dnn-coding/}.
	\end{abstract}
	
	
	\IEEEpeerreviewmaketitle
	
	\section{Introduction}
\label{sec:introduction}

In recent years, Internet traffic has been dominated by a wide range of applications involving video, including video on demand (VOD), live streaming, ultra-low latency real-time communications, {\it etc.}. With ever increasing demands in resolution ({\it e.g.}, 4K, 8K, gigapixel~\cite{brady2012multiscale}, high speed~\cite{cheng2020dual}), and fidelity, ({\it e.g.}, high dynamic range~\cite{dufaux2016high}, and higher bit precision or bit depth~\cite{winken2007bit}), more efficient video compression is imperative for content transmission and storage, by which networked video services can be successfully deployed.
Fundamentally, video compression systems devise appropriate algorithms to minimize the end-to-end reconstruction distortion (or maximize the quality of experience (QoE)), under a given bit rate budget. This is a classical rate-distortion (R-D) optimization problem. 
In the past, the majority of effort had been focused on the development and standardization of video coding tools for optimized R-D performance, such as the intra/inter prediction, transform, entropy coding, {\it etc.}, resulting in a number of popular standards and recommendation specifications ({\it e.g.}, ISO/IEC MPEG series~\cite{tudor1995mpeg, haskell1996digital, sikora1997mpeg, li2001overview, wiegand2003overview, sullivan2012overview, sze2014high}, ITU-T H.26x series~\cite{wiegand2003overview, sullivan2004h, vetro2011overview, sullivan2012overview, sze2014high}, AVS series~\cite{yu2009overview, ma2013overview, zhang2019recent}, as well as the AV1~\cite{chen2018overview, han2020technical} from the Alliance of Open Media (AOM)\cite{AOM}). 
All these standards have been widely deployed in the market and enabled advanced and high-performing services to both enterprises and consumers. They have been adopted to cover all major video scenarios from VOD, to live streaming, to ultra-low latency interactive real-time communications, used for applications such as telemedicine, distance learning, video conferencing, broadcasting, e-commerce, online gaming, short video platforms, \textit{etc}. 
Meanwhile, the system R-D efficiency can also be improved from pre-processing and post-processing, individually and jointly, for content adaptive encoding (CAE). Notable examples include saliency detection for subsequent region-wise quantization control, and   adaptive filters to alleviate compression distortions~\cite{netflix-pertile,shoham2019,lin-icip2015}.

{In this article, we therefore consider {\it pre-processing}, {\it coding}, and {\it post-processing} as three basic functional blocks of an end-to-end video compression system, and optimize them to provide compact and high-quality representation of input original video.
\begin{itemize}
    \item The ``coding'' block is the core unit that converts raw pixels or pixel blocks into binary bits presentation. Over the past decades, the ``coding'' R-D efficiency has been gradually improved by introducing more advanced tools to better exploit spatial, temporal, and statistical redundancy~\cite{sullivan2005video}. Nevertheless, this process inevitably incurs compression artifacts, such as blockiness and ringing, due to the R-D trade-off, especially at low bit rates.
    \item The ``post-processing'' block is introduced to alleviate visually perceptible impairments produced as byproducts of coding. Post-processing mostly relies on the designated adaptive filters to enhance the reconstructed video quality or QoE. Such ``post-processing'' filters can also be embedded into the ``coding'' loop to jointly improve reconstruction quality and R-D efficiency, {\it e.g.}, in-loop deblocking~\cite{norkin2012hevc} and sample adaptive offset (SAO)~\cite{fu2012sample}; 
    \item The ``pre-processing" block exploits the discriminative content preference of the human visual system (HVS), caused by the non-linear response and frequency selectivity ({\it e.g.}, masking) of visual neurons in the visual pathway.  Pre-processing can extract content semantics ({\it e.g.}, saliency, object instance) to improve the psychovisual performance of the ``coding'' block, for example, by allocating unequal qualities (UEQ) across different areas according to pre-processed cues~\cite{gupta2013visual}. 
    \footnote{Although adaptive filters can also be used in pre-processing for pre-filtering, {\it e.g.}, denoising, motion deblurring, contrast enhancement, edge detection, {\it etc.}, our primary focus in this work will be on semantic content understanding for subsequent intelligent ``coding''.}
    \end{itemize}  }


Building upon the advancements in  deep neural networks (DNN), numerous recently-created video processing algorithms have been greatly improved to achieve superior performance, mostly leveraging the powerful nonlinear representation capacity of DNNs. At the same time, we have also witnessed an explosive growth in the invention of DNN-based techniques for video compression from both academic research and industrial practices. 
{For example, DNN-based filtering in post-processing was extensively studied when developing the VVC standard under the joint task force of ISO/IEC and ITU-T experts over the past three years. More recently, the standard committee issued a Call-for-Evidence (CfE)~\cite{jvet-ahg_on_nncoding, jvet-report_of_ahg11} to encourage the exploration of deep learning-based video coding solutions beyond VVC.}

\begin{figure*}[!t]
	\centering
	\centerline{\includegraphics[width=\linewidth]{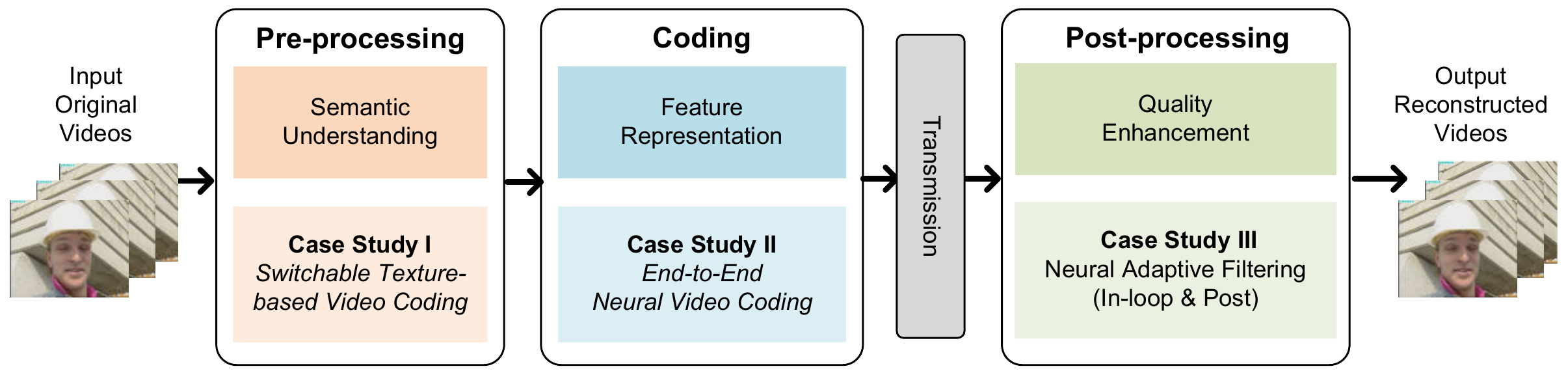}}
	\caption{\textbf{Topic Outline.} This article reviews DNN-based techniques used in pre-processing, coding, and post-processing of a practical video compression system. The ``pre-processing'' module leverages content semantics ({\it e.g.}, texture) to guide video coding, followed by the ``coding'' step to represent the video content using more compact spatio-temporal features. Finally, quality enhancement is applied in ``post-processing'' to improve reconstruction quality by alleviating processing artifacts. Companion case studies are respectively offered to showcase the potential of DNN algorithms in video compression.}
	\label{fig:sec1_introduction_contribution}
\end{figure*}

In this article, we discuss recent advances in {\it pre-processing}, {\it coding}, and {\it post-processing}, with particular emphasis on the use of DNN-based approaches for efficient video compression. We aim to provide a comprehensive overview to bring readers up to date on recent advances in this emerging field. We also suggest promising directions for further exploration.
As summarized in Fig.~\ref{fig:sec1_introduction_contribution}, we first dive into video pre-processing, emphasizing the analysis and application of content semantics, {\it e.g.}, saliency, object, texture characteristics, {\it etc.}, to video encoding. We then discuss recently-developed DNN-based video coding techniques for both modularized coding tool development and end-to-end fully learned framework exploration. Finally, we provide an overview of the adaptive filters that can be either embedded in codec loop, or placed as a post enhancement to improve final reconstruction. We also present three case studies, including 1) \emph{switchable texture-based video coding} in pre-processing; 2) \emph{end-to-end neural video coding}; and 3) \emph{efficient neural filtering}, to provide examples the potential of DNNs to improve both subjective and objective efficiency over traditional video compression methodologies.

The remainder of the article is organized as follows: From Section~\ref{sec:review_pre_processing} to~\ref{sec:review_dnn_post_processing}, we extensively review the advances in respective pre-processing, coding, and post-processing. Traditional methodologies are first briefly summarized, and then DNN-based approaches are discussed in detail.
As in the case studies, we propose three neural approaches in Section~\ref{sec:proposed_pre_processing}, \ref{sec:proposed_end_to_end_coding}, and \ref{sec:proposed_post_processing}, respectively. Regarding pre-processing, we develop a CNN based texture analysis/synthesis scheme for AV1 codec. For video compression, an end-to-end neural coding framework is developed. In our discussion of post-processing,we present different neural methods for in-loop and post filtering  that can enhance the quality of reconstructed frames. Section~\ref{sec:conclusion} summarizes this work and discusses open challenges and future research directions. For your convenience, Table~\ref{tab:abber} provides an overview of abbreviations and acronyms that are frequently used throughout this paper.

\begin{table}[t]
\centering
\caption{Abbreviations and Annotations}
\label{tab:abber}
\begin{tabular}{c|c}
\hline
    Abbreviation & Description  \\
    \hline
    AE & AutoEncoder\\
    CNN & Convolutional Neural Network\\
    CONV & Convolution\\
    ConvLSTM & Convolutional LSTM\\
    DNN & Deep Neural Network\\
    FCN & Fully-Connected Network\\
    GAN & Generative Adversarial Network\\
    LSTM & Long Short-Term Memory \\
    RNN & Recurrent Neural Network\\
    VAE & Variational AutoEncoder\\
    \hline
    BD-PSNR & Bj\o ntegaard Delta PSNR\\
    BD-Rate & Bj\o ntegaard Delta Rate\\
    GOP & Group of Pictures \\
    MS-SSIM & Multiscale SSIM\\
    MSE & Mean Squared Error\\
    PSNR & Peak Signal-to-Noise Ratio\\
    QP & Quantizatin Parameter\\
    QoE & Quality of Experience\\
    SSIM & Structural Similarity Index\\
    UEQ & UnEqual Quality\\
    VMAF & Video Multi-Method Assessment Fusion\\
    \hline
    AV1 & AOMedia Video 1 \\
    AVS & Audio Video Standard \\
    H.264/AVC & H.264/Advanced Video Coding\\
    H.265/HEVC & H.265/High-Efficiency Video Coding\\
    VVC & Versatile Video Coding\\
    \hline
    AOM & Alliance of Open Media\\
    MPEG & Moving Picture Experts Group\\
    \hline
\end{tabular}
\end{table}
	
\section{Overview of DNN-based Video Pre-processing}
\label{sec:review_pre_processing}

{Pre-processing techniques are generally applied prior to the video coding block, with the objective of guiding the video encoder to remove psychovisual redundancy and to maintain or improve visual quality, while simultaneously lowering bit rate consumption.
One category of pre-processing techniques is the execution of pre-filtering operations. Recently, a number of deep learning-based pre-filtering approaches have been adopted for targeted coding optimization. These include  denoising~\cite{zhang2017beyond,tian2018deep}, motion deblurring~\cite{chakrabarti2016neural, koh2020single}, contrast enhancement~\cite{zhu2020learning}, edge detection~\cite{guan2018edge, xu2015deep},  {\it etc}. 
Another important topic area is closely related to the analysis of video content semantics, {\it e.g.}, object instance, saliency attention, texture distribution, {\it etc.}, and its application to intelligent video coding.}
For the sake of simplicity, we refer to this group of techniques as ``pre-processing'' for the remainder of this paper. In our discussion below, we also limit our focus to saliency-based and analysis/synthesis-based approaches.


\subsection{Saliency-Based Video Pre-processing}

\subsubsection{Saliency Prediction}

{\textit{Saliency} is the quality of being particularly noticeable or important. Thus, the {\it salient area}  refers to region of an image that predominantly attracts the attention of subjects. This concept corresponds closely to the highly discriminative and selective behaviour displayed in visual neuronal processing ~\cite{zhaoping2019new,chen2019contribution}. Content feature extraction, activation, suppression and aggregation also occur in the visual pathway~\cite{schwartz2001natural}. }

{Earlier attempts to predict saliency typically utilized handcrafted image features, such as color, intensity, and orientation contrast~\cite{itti1998model}; motion contrast~\cite{itti2004automatic}; camera motion~\cite{nguyen2013static}, {\it etc.}, to predict saliency. 

Later on, DNN-based semantic-level features
were extensively investigated for both image content~\cite{vig2014large, liu2015predicting, li2016deep, liu2016dhsnet, hou2017deeply, yan2017accurate, xu2018personalized} and video sequences~\cite{bazzani2016recurrent, bak2017spatio, sun2018sg, jiang2018deepvs, wang2018deep, cong2019video, min2019tased}. Among these features, image saliency prediction only exploits spatial information, while video saliency prediction often relies on  spatial and temporal attributes jointly. One typical example of video saliency is a moving object that incurs spatio-temporal dynamics over time, and is therefore more likely to attract users' attention.   For example, 
Bazzani~\emph{et al.}~\cite{bazzani2016recurrent} modeled the spatial relations in videos using 3D convolutional features and the temporal consistency with a convolutional long short-term memory (LSTM) network. 
Bak~\emph{et al.}~\cite{bak2017spatio} applied a two-stream network  that exploited different fusion mechanisms to effectively integrate spatial and temporal information.
Sun~\emph{et al.}~\cite{sun2018sg} proposed a step-gained FCN to combine the time-domain memory information and space-domain motion components. 
Jiang~\emph{et al.}~\cite{jiang2018deepvs} developed an object-to-motion CNN that was applied together with a LSTM network. All of these efforts to efficiently predict video saliency leveraged spatio-temporal attributes.
More details regarding the spatio-temporal saliency models for video content  can be found in~\cite{wang2019revisiting}.}

\subsubsection{Salient Object}
{One special example of image saliency involved the {\it object instance} in a visual scene, specifically, the moving object in videos. A simple yet effective solution to the problem of predicting image saliency in this case involved segmenting foreground objects and background components. }

{The segmentation of foreground objects and background components has mainly relied on foreground extraction or background subtraction. For example, motion information has frequently been used to mask out foreground objects~\cite{1221642, 1334528, zhu2010motion, zhang2017flow, guo2019foreground}. 

Recently, both CNN and foreground attentive neural network (FANN) models have been developed to perform foreground segmentation~\cite{shahbaz2019convolutional,zhou2019discriminative}. In addition to conventional Gaussian mixture model-based background subtraction, recent explorations have also shown that CNN models could be effectively used for the same purpose~\cite{babaee2017deep,liang2018deep}.}
{To address these separated foreground objects and background attributes, Zhang \emph{et al.}~\cite{zhang2012efficient} introduced a new background mode to more compactly represent background information with better R-D efficiency. To the best of our knowledge, 
such foreground object/background segmentation has been mostly applied in video surveillance applications, where the visual scene lends itself to easier separation.}


\subsubsection{Video Compression with UEQ Scales}
{Recalling that saliency or object refers to more visually attentive areas. It is straightforward to apply UEQ setting in a video encoder, where light compression is used to encode the saliency area, while heavy compression is used elsewhere. Use of this technique often results in a lower level of total bit rate consumption without compromising QoE. 

For example, Hadi \emph{et al.}~\cite{hadizadeh2013saliency} extended the well-known Itti-Koch-Niebur (IKN) model to estimate saliency in the DCT domain, also considering camera motion. In addition, saliency-driven distortion was also introduced to accurately capture the salient characteristics, in order to improve R-D optimization in H.265/HEVC.
Li \emph{et al.}~\cite{li2014saliency} suggested using graph-based visual saliency to adapt the quantizations in H.265/HEVC, to reduce total bits consumption.
Similarly, Ku~\emph{et al.}~\cite{ku2019bit} applied saliency-weighted Coding Tree Unit (CTU)-level bit allocation, where the CTU-aligned saliency weights were determined via low-level feature fusion.}

{The aforementioned methodologies rely on traditional handcrafted saliency prediction algorithms. As DNN-based saliency algorithms have demonstrated superior performance, we can safely assume that their application to video coding will lead to better compression efficiency. For example,
Zhu~\emph{et al.}~\cite{zhu2018spatiotemporal} adopted a spatio-temporal saliency model to accurately control the QP in an encoder whose spatial saliency was generated using a 10-layer CNN, and whose temporal saliency was calculated assuming the 2D motion model (resulting in an average of 0.24 BD-PSNR gains over H.265/HEVC reference model (version HM16.8)).
Performance improvement due to fine-grained quantization adaptation was reported using an open-source x264 encoder~\cite{lyudvichenko2019improving}. This was accomplished by jointly examining the input video frame and associated saliency maps. These saliency maps were generated by utilizing three CNN models suggested in~\cite{jiang2018deepvs,wang2019revisiting,cornia2018predicting}.
Up to 25\% bit rate reduction was reported when distortion was measured using the edge-weighted SSIM (EW-SSIM).
Similarly, Sun~\emph{et al.}~\cite{sun2020content} implemented a saliency-driven CTU-level adaptive bit rate control, where the static saliency map of each frame was extracted using a DNN model and dynamic saliency region when it was tracked using a moving object segmentation algorithm. Experiment results revealed that the PSNR of salient regions was improved by 1.85 dB on average.
} 

{Though saliency-based pre-processing is mainly driven by psychovisual studies, it heavily relies on saliency detection to perform UEQ-based adaptive quantization with a lower rate of bit consumption but visually identical reconstruction. On the other hand, visual selectivity behaviour is closely associated with video content distribution ({\it e.g.}, frequency response), leading to perceptually unequal preference. Thus, it is highly expected that such content semantics-induced discriminative features can be utilized to improve the system efficiency when integrated into the video encoder. To this end, we will discuss the analysis/synthesis-based approach for pre-processing in the next section.}

\subsection{Analysis/Synthesis Based Pre-processing}
\label{subsec:texture_based_preprocessing}
{Since most videos are consumed by human vision, subjective perception of HVS is the {\it best} way to evaluate quality. However, it is quite difficult to devise a profoundly accurate mathematical HVS model in actual video encoder for rate and perceptual quality optimization, due to the complicated and unclear information processing that occurs in the human visual pathway. Instead,  many pioneering psychovisual studies have suggested that neuronal response to compound stimuli is highly nonlinear~\cite{carandini2005we,kremkow2014neuronal,ukita2019characterisation,neri2009nonlinear,heeger1992normalization,priebe2012mechanisms,carandini2012normalization,turner2016synaptic} within the receptive field. This leads to well-known visual behaviors, such as frequency selectivity, masking, {\it etc.}, where such stimuli are closely related to the content texture characteristics.  Intuitively, video scenes can be broken down into areas that are either ``perceptually significant'' ({\it e.g.}, measured in an MSE sense) or ``perceptually insignificant''. For ``perceptually insignificant" regions,  users will not perceive compression or processing impairments without a side-by-side comparison with the original sample. This is because the HVS gains semantic understanding by viewing content as a whole, instead of interpreting  texture details pixel-by-pixel~\cite{DOSHKOV2014197}. This notable effect of the HVS is also referred to as ``masking,'' where visually insignificant information, {\it e.g.}, perceptually insignificant pixels, will be noticeably suppressed.}

\begin{figure}[t]
  \centering
  \includegraphics[width=0.80\linewidth]{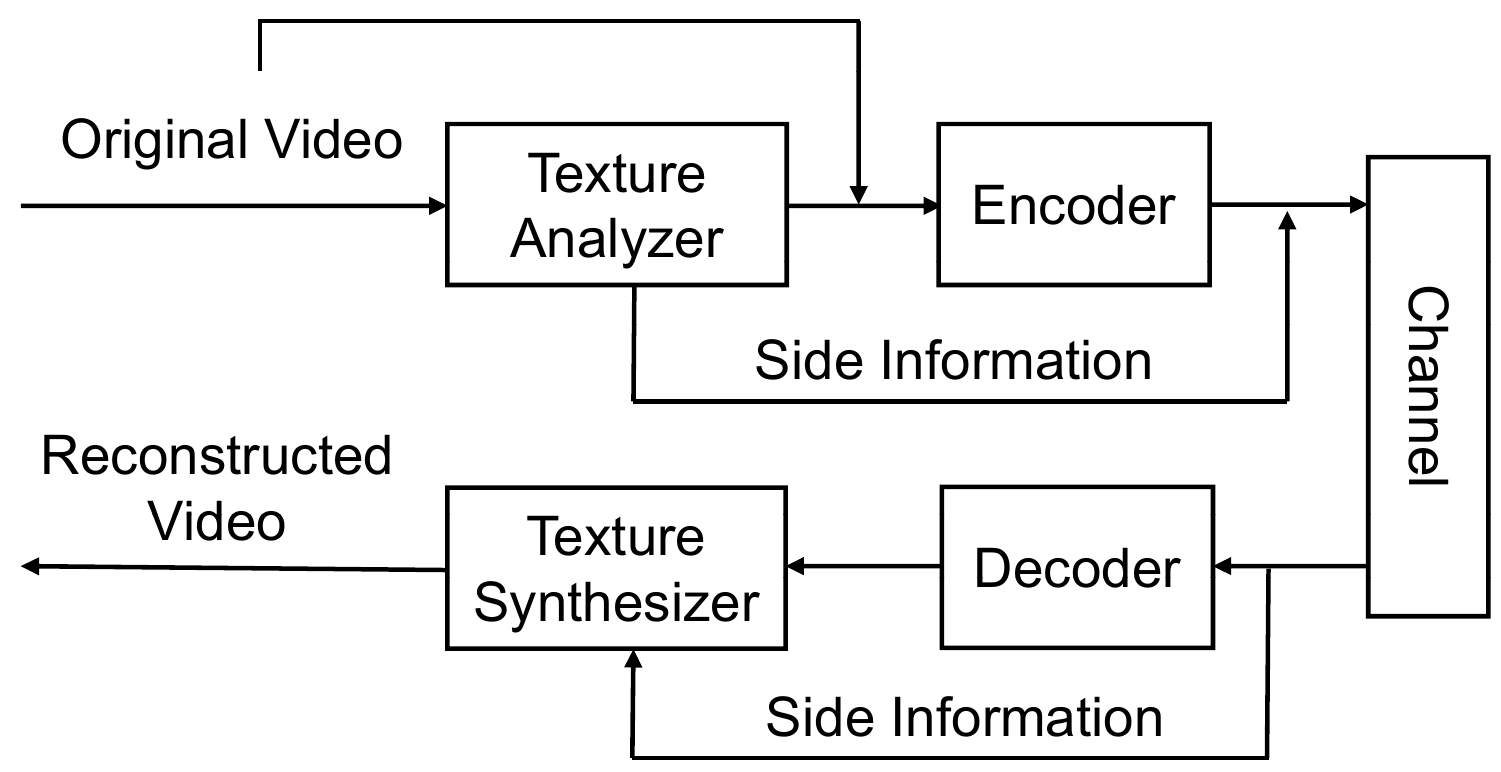}
\caption{{\bf Texture Coding System.} {A general framework of analysis/synthesis based video coding.}}
\label{fig:sec2_texture_analysis_overview}
\end{figure}

In practice, we can first analyze the texture characteristics of original video content in the pre-processing step, {\it e.g.}, {\it Texture Analyzer} in Fig.~\ref{fig:sec2_texture_analysis_overview}, in order to sort textures by their significance. Subsequently, we can use any standard compliant video encoder to encode the perceptually significant areas as the main bitstream payload, and apply a statistical model to represent the perceptually insignificant textures with model parameters encapsulated as side information. Finally, we can use decoded areas and parsed textures to jointly synthesize the reconstructed sequences in {\it Texture Synthesizer}. This type of texture modeling makes good use of statistical and psychovisual representation jointly, generally requiring fewer bits, despite yielding visually identical sensation, compared to the traditional hybrid ``prediction+residual'' method\footnote{A comprehensive survey of texture analysis/synthesis based video coding technologies can be found in~\cite{NDJIKINYA2012579}. }. Therefore, texture analysis and synthesis play a vital role for subsequent video coding. We will discuss related techniques below.

\subsubsection{Texture Analysis}
{Early developments in texture analysis and representation can be categorized into  {\it filter-based} or {\it statistical modeling-based} approaches.}
{Gabor filter is one typical example of a filter-based approach, by which the input image is convoluted with nonlinear activation for the derivation of corresponding texture representation~\cite{jain1990unsupervised,bovik1990multichannel}. At the same time, in order to identify static and dynamic textures for video content,  Thakur {\it et al.}~\cite{7314212} utilized the 2D dual tree complex wavelet transform and steerable pyramid transform~\cite{portilla2000parametric}, respectively. To accurately capture the temporal variations in video, Bansal {\it et al.}~\cite{bansal2013} again suggested the use of optic flow  for dynamic texture indication and later synthesis, where optical flow could be generated using temporal filtering.}
{Leveraging statistical models such as the Markovian random field (MRF)~\cite{cross1983markov, chellappa1985classification} is an alternative way to analyze and represent texture. For efficient texture description, statistical modeling such as this was then extended using handcrafted local features, {\it e.g.}, the scale invariant feature transform (SIFT)~\cite{lowe2004distinctive}, speeded up robust features (SURF)~\cite{bay2006surf}, and local binary patterns (LBP)~\cite{ojala2002multiresolution}}

{Recently, stacked DNNs have demonstrated their superior efficiency in many computer vision tasks, This efficiency is mainly due to the powerful capacity of DNN features to be used for video content representation.  The most straightforward scheme directly extracted features from the FC6 or FC7 layer of AlexNet~\cite{krizhevsky2012imagenet} for texture representation. Furthermore, Cimpoi {\it et al.}~\cite{cimpoi2015deep} demonstrated that Fisher vectorized~\cite{perronnin2010improving} CNN features was a decent texture descriptor candidate.}

\subsubsection{Texture Synthesis}
{Texture synthesis reverse-engineers the analysis in pre-processing to restore pixels accordingly. It generally includes both non-parametric and parametric methods.}
{For non-parametric synthesis, texture patches are usually resampled from reference images~\cite{efros1999texture,wei2000fast,ashikhmin2001synthesizing}. In contrast, the parametric method utilized statistical models to reconstruct the texture regions by jointly optimizing  observation outcomes from the model and model itself~\cite{derin1987modeling, heeger1995pyramid, portilla2000parametric}.}

{DNN-based solutions exhibit great potential for texture synthesis applications. One notable example demonstrating this potential used a pre-trained image classification-based CNN model to generate texture patches~\cite{gatys2015texture}. Li \etal~\cite{li2016precomputed}, then demonstrated that a Markovian GAN-based texture synthesis could offer remarkable quality improvement.}

To briefly summarize, earlier ``texture analysis/synthesis'' approaches often relied on handcrafted models, as well as corresponding parameters. While they have shown good performance to some extent for a set of test videos, it is usually very difficult to generalize them to large-scale video datasets without fine-tuning parameters further. On the other hand, related neuroscience studies propose a  broader definition of texture which is more closely related to perceptual sensation, although existing mathematical or data-driven texture representations attempt to fully fulfill such perceptual motives. Furthermore, recent DNN-based schemes present a promising perspective. However, the complexity of these schemes has not yet been appropriately exploited. So, in Section~\ref{sec:proposed_pre_processing}, we will reveal a CNN-based pixel-level texture analysis approach to segment perceptually insignificant texture areas in a frame for compression and later synthesis. To model the textures both spatially and temporally, we introduce a new coding mode called the ``switchable texture mode" that is determined at group of pictures (GoP) level according to the bit rate saving.

\section{Overview of DNN-based Video Coding}
\label{sec:review_dnn_based_coding}
A number of investigations have shown that DNNs can be used for efficient image/video coding~\cite{chen2017deepcoder,balle2016end,liu2020deep,ma2019image}. This topic has attracted extensive attention in recent years, demonstrating its potential to enhance the conventional system with better R-D performance.  

{There are three major directions currently under investigation. One is resolution resampling-based video coding, by which the input videos are first down-sampled prior to being encoded, and the reconstructed videos are up-sampled or super-resolved to the same  resolution as the input~\cite{li2018convolutional,jiang2017end, afonso2018video, lin2018convolutional}. This category generally develops up-scaling or super-resolution algorithms on top of standard video codecs.} The second direction under investigation is  modularized  neural video coding (MOD-NVC), which has attempted to improve individual coding tools in traditional hybrid coding framework using learning-based solutions. The third direction is end-to-end neural video coding (E2E-NVC), which fully leverages the stacked neural networks to compactly represent input image/video in an end-to-end learning manner. In the following sections, we will primarily review the latter two cases, since the first one has been extensively discussed in many other studies~\cite{yang2019deep}.

\subsection{Modularized Neural Video Coding (MOD-NVC)}
\label{module_based}
The MOD-NVC has inherited the traditional hybrid coding framework within which handcrafted tools are refined or replaced using learned solutions. The general assumption is that existing rule-based coding tools can be further improved via a data-driven approach that leverages powerful DNNs to learn robust and efficient mapping functions for more compact content representation. 
Two great articles have comprehensively reviewed relevant studies in this direction~\cite{ma2019image, liu2020deep}.
We briefly introduce key techniques in intra/inter prediction, quantization, and entropy coding.
Though in-loop filtering is another important piece in the ``coding'' block, due to its similarities with post filtering, we have chosen to review it in quality enhancement-aimed ``post-processing'' for the sake of creating a more cohesive presentation.

\subsubsection{Intra Prediction}
{Video frame content presents highly correlated distribution across neighboring samples spatially. Thus, block redundancy can be effectively exploited using causal neighbors.
In the meantime, due to the presence of local structural dynamics, block pixels can be better represented from a variety of angular directed prediction.}

{In conventional standards, such as the H.264/AVC, H.265/HEVC, or even emerging VVC, specific prediction rules are carefully designated to use weighted neighbors for respective angular directions. From the H.264/AVC to recent VVC, intra coding efficiency has been gradually improved by allowing more fine-grained angular directions and flexible block size/partitions. In practice, an optimal coding mode is often determined by R-D optimization. }

{One would intuitively expect that coding performance can be further improved if  better predictions can be produced. Therefore, there have been a number of attempts to leverage the powerful capacity of stacked DNNs for better intra predictor generation, including the CNN-based predictor refinement suggested in~\cite{cui2018convolutional} to reduce prediction residual, additional learned mode trained using FCN models reported in~\cite{li2018fully,pfaff2018neural}, using RNNs in~\cite{hu2019progressive}, using CNNs in~\cite{li2018convolutional}, or even using GANs in~\cite{jin2020video}, {\it etc.} These approaches have actively utilized the neighbor pixels or blocks, and/or other context information ({\it e.g.}, mode) if applicable, in order to accurately represent the local structures for better prediction. Many of these approaches have reported more than 3\% BD-Rate gains against the popular H.265/HEVC reference model. These examples demonstrate the efficiency of DNNs in intra prediction. }

\subsubsection{Inter Prediction}
{In addition to the spatial intra prediction, temporal correlations have also been exploited via {\it inter prediction,} by which previously reconstructed frames are utilized to generate inter predictor for compensation using displaced motion vectors.}

{Temporal prediction can be enhanced using references with higher fidelity, and more fine-grained motion compensation. For example, fractional-pel interpolation is usually deployed to improve prediction accuracy~\cite{girod1993motion}. On the other hand, motion compensation with flexible block partitions is another major contributor to inter coding efficiency.}

{Similarly, earlier attempts have been made to utilize DNNs solutions for better inter coding. For instance, CNN-based  interpolations were studied in~\cite{yan2017convolutional,zhang2017learning,liu2018one}
to improve the half-pel samples. Besides, an additional virtual reference could be generated using CNN models for improved R-D decision in~\cite{zhao2019enhanced}. Xia \etal~\cite{xia2019deep} further extended this approach using multiscale CNNs to create an additional reference closer to the current frame by which accurate pixel-wise motion representation could be used.  Furthermore, conventional references could also be enhanced using DNNs to refine the compensation~\cite{huo2018convolutional}.}

\subsubsection{Quantization and Entropy Coding}
{Quantization and entropy coding are used to remove statistical redundancy. 
Scalar quantization is typically implemented in video encoders to remove insensitive high-frequency components, without losing the perceptual quality, while saving the bit rate. Recently, a three-layer DNN was developed to predict the local visibility threshold $C_{T}$ for each CTU, by which more accurate quantization could be achieved via the connection between $C_{T}$ and actual quantization stepsize. This development led to noticeable R-D improvement, {\it e.g.}, upto 11\% as reported in~\cite{alam2015perceptual}. }

{Context-adaptive binary arithmetic coding (CABAC) and its variants are techniques that are widely adopted to encode binarized symbols. The efficiency of CABAC is heavily reliant on the accuracy of probability estimation in different contexts. Since the H.264/AVC, handcrafted probability transfer functions (developed through exhaustive simulations, and typically implemented using look-up tables) were utilized. }
{In~\cite{pfaff2018neural} and~\cite{song2017neural}, the authors demonstrated that a combined FCN and CNN model could be used to predict intra mode probability for better entropy coding. Another example of a combined FCN and CNN model was presented in~\cite{puri2017cnn} to accurately encode transform indexes via stacked CNNs. And likewise, in~\cite{ma2018convolutional}, intra DC coefficient probability could be also estimated using DNNs for better performance. }

{All of these explorations have reported positive R-D gains when incorporating DNNs in traditional hybrid coding frameworks. A companion H.265/HEVC-based software model is also offered by Liu \etal~\cite{liu2020deep}, to advance the potential for society to further pursue this line of exploration. However, integrating DNN-based tools could exponentially increase both the computational and space complexity. Therefore, creating harmony between learning-based 
and conventional rule-based tools under the same framework requires further investigation. It is also worth noting that an alternative approach is currently being explored in parallel. In this approach, researchers suggest using an end-to-end neural video coding (E2E-NVC) framework to drive the raw video content representation via layered feature extraction, activation, suppression, and aggregation, mostly in a supervised learning fashion, instead of refining individual coding tools. }

\subsection{End-to-End Neural Video Coding (E2E-NVC)}
\label{end_to_end}

{Representing raw video pixels as compactly as possible by massively exploiting its spatio-temporal and statistical correlations is the fundamental problem of lossy video coding. Over decades, traditional hybrid coding frameworks have utilized pixel-domain intra/inter prediction, transform, entropy coding, {\it etc.}, to fulfill this purpose. Each coding tool is extensively examined under a specific codec structure to carefully justify the trade-off between R-D efficiency and complexity. This process led to the creation of well-known international or industry standards, such as the H.264/AVC, H.265/HEVC, AV1, {\it etc.}}

{On the other hand, DNNs have demonstrated a powerful capacity for video spatio-temporal feature representation for vision tasks, such as object segmentation, tracking, {\it etc.} This naturally raises the question of whether it is possible to encode those spatio-temporal features in a compact format for efficient lossy compression. }

Recently, we have witnessed the growth of video coding technologies that rely completely on end-to-end supervised learning. Most learned schemes still closely follow the conventional intra/inter frame definition by which different algorithms are investigated to efficiently represent the intra spatial textures, inter motion, and the inter residuals (if applicable)~\cite{chen2017deepcoder,lu2019dvc,rippel2019learned,liu2020learned}. 
{Raw video frames are fed into stacked DNNs to extract, activate, and aggregate appropriate compact features (at the bottleneck layer) for quantization and entropy coding. Similarly, R-D optimization is also facilitated to balance the rate and distortion trade-off. In the following paragraphs, we will briefly  review the aforementioned key components.}

\subsubsection{Nonlinear Transform and Quantization}

The autoencoder or variational autoencoder (VAE) architectures are typically used to transform the intra texture or inter residual into compressible features.

For example, Toderic \etal~\cite{toderici2015variable} first applied fully-connected {\it recurrent autoencoders} for variable-rate thumbnail image compression. Their work was then improved in ~\cite{toderici2017full,johnston2018improved} with the support of full-resolution image, unequal bit allocation, {\it etc.} Variable bit rate is intrinsically enabled by these recurrent structures. The recurrent autoencoders, however, suffer from higher computational complexity at higher bit rates, because more recurrent processing is desired. Alternatively, {\it convolutional autoencoders} have been extensively studied in past years, where different bit rates are adapted by setting a variety of $\lambda$s to optimize the R-D trade-off.  Note that different network models may be required for individual bit rates, making hardware implementation challenging, ({\it e.g.}, model switch from one bit rate to another). Recently, conditional convolution~\cite{choi2019variable} and scaling factor~\cite{chen2019neural} were proposed to enable variable-rate compression using a single or very limited network model without noticeable coding efficiency loss, which makes the convolutional autoencoders more attractive for practical applications.

To generate a more compact feature representation, Balle \etal~\cite{balle2016end} suggested replacing the traditional nonlinear activation, {\it e.g.}, ReLU, using generalized divisive normalization (GDN) that is theoretically proven to be more consistent with human visual perception. A subsequent study~\cite{balle2018efficient} revealed that GDN outperformed other nonlinear rectifiers, such as ReLU, leakyReLU, and tanh, in compression tasks. Several follow-up studies~\cite{lee2018context,klopp2018learning} directly applied GDN in their networks for compression exploration.

Quantization is a non-differentiable operation, basically converting arbitrary elements into symbols with a limited alphabet for efficient entropy coding in compression. Quantization must be derivable in the end-to-end learning framework for back propagation. A number of methods, such as adding uniform noise~\cite{balle2016end}, stochastic rounding~\cite{toderici2015variable} and soft-to-hard vector quantization~\cite{mentzer2018conditional}, were developed to approximate a continuous distribution for differentiation.

\subsubsection{Motion Representation}

Chen \etal~\cite{chen2017deepcoder} developed the DeepCoder where a simple convolutional autoencoder was applied for both intra and residual coding at fixed 32$\times$32 blocks, and block-based motion estimation in traditional video coding was re-used for temporal compensation. Lu \etal~\cite{Lu_2019_CVPR} introduced the optical flow for motion representation in their DVC work, which, together with the intra coding in~\cite{balle2018variational}, demonstrated similar performance compared with the H.265/HEVC. However, coding efficiency suffered from a sharp loss at low bit rates. Liu \etal~\cite{liu2019non} extended their non-local attention optimized image compression (NLAIC) for intra and residual encoding, and applied second-order flow-to-flow prediction for more compact motion representation, showing consistent rate-distortion gains across different contents and bit rates. 

Motion can also be implicitly inferred via temporal interpolation. For example, Wu \etal~\cite{wu2018vcii} applied RNN-based frame interpolation. Together with the residual compensation, RNN-based frame interpolation offered comparable performance to the H.264/AVC. Djelouah \etal~\cite{djelouah2019neural} furthered interpolation-based video coding by utilizing advanced optical flow estimation and feature domain residual coding. However, temporal interpolation usually led to an inevitable structural coding delay.

Another interesting exploration made by Ripple \emph{et al.} in~\cite{rippel2019learned} was to jointly encode motion flow and residual using compound features, where a recurrent state was embedded to aggregate multi-frame information for efficient flow generation and residual coding.

\subsubsection{R-D Optimization}

Li \etal~\cite{li2017learning} utilized a separate three-layer CNN to generate an importance map for spatial-complexity-based adaptive bit allocation, leading to noticeable subjective quality improvement. Mentzer \etal~\cite{mentzer2018conditional} further utilized the masked bottleneck layer to unequally weight features at different spatial locations. Such importance map embedding is a straightforward approach to end-to-end training.
Importance derivation was later improved with the non-local attention~\cite{wang2018non} mechanism to efficiently and implicitly capture both global and local significance for better compression performance~\cite{chen2019neural}.


Probabilistic models play a vital role in data compression. Assuming the Gaussian distribution for feature elements, Balle \etal~\cite{balle2018variational} utilized hyper priors to estimate the parameters of Gaussian scale model (GSM) for latent features. Later Hu \etal~\cite{hu2020coarse} used hierarchical hyper priors (coarse-to-fine) to improve the entropy models in multiscale representations.  Minnen \etal~\cite{minnen2018joint} improved the context modeling using joint autoregressive spatial neighbors and hyper priors based on the Gaussian mixture model (GMM). Autoregressive spatial priors were commonly fused by PixelCNNs or PixelRNNs~\cite{oord2016pixel}. Reed \etal~\cite{reed2017parallel} further introduced multiscale PixelCNNs, yielding competitive density estimation and great boost in speed ({\it e.g.}, from $O(N)$ to $O(\log N)$). Prior aggregation was later extended from 2D architectures to 3D PixelCNNs~\cite{mentzer2018conditional}. Channel-wise weights sharing-based 3D implementations could greatly reduce network parameters without performance loss. A parallel 3D PixelCNNs for practical decoding is presented in Chen \etal~\cite{chen2019neural}. Previous methods accumulated all the priors to estimate the probability based on a single GMM assumption for each element. Recent studies have shown that weighted GMMs can further improve coding efficiency in~\cite{cheng2020learned,lee2019end}.

Pixel-error, such as MSE, was one of the most popular loss functions used. Concurrently, SSIM (or MS-SSIM) was also adopted because of its greater consistency with visual perception. Simulations  revealed that SSIM-based loss can improve reconstruction quality, especially at low bit rates.
Towards the perceptual-optimized encoding, perceptual losses that were measured by adversarial loss~\cite{rippel2017real,huang2019extreme,agustsson2019generative} and VGG loss~\cite{liu2018deep} were embedded in learning to produce visually appealing results.

Though E2E-NVC is still in its infancy, its fast growing R-D efficiency holds a great deal of promise. This is especially true, given that we can expect neural processors to be deployed massively in the near future~\cite{hennessy2019new}.

	\section{Overview of DNN-based Post-processing}
\label{sec:review_dnn_post_processing}
{Compression artifacts are inevitably present in both traditional hybrid coding frameworks and learned compression approaches, {\it e.g.}, blockiness, ringing, cartoonishness, {\it etc.}, severely impairing visual sensation and QoE. Thus, quality enhancement filters are often applied as a post-filtering step or in-loop module to alleviate compression distortions. Towards this goal, adaptive filters are usually developed to minimize the error between original and distorted samples.}

\subsection{In-loop Filtering}
\label{subsec:in-loop filtering}

Existing video standards are mainly utilizing the in-loop filters to improve the subjective quality of reconstruction, and also to offer better R-D efficiency due to enhanced references. Examples include deblocking~\cite{norkin2012hevc}, sample adaptive offset (SAO)~\cite{fu2012sample},  constrained directional enhancement filter (CDEF)~\cite{midtskogen2018av1}, loop-restoration (LR)~\cite{mukherjee2017switchable}, adaptive loop filter (ALF)~\cite{tsai2013adaptive}, {\it etc.}

Recently, numerous CNN models have been developed for in-loop filtering via a data-driven approach to learn the mapping functions. It is worth pointing out that prediction relationships must be carefully examined when designing in-loop filters, due to the frame referencing structure and potential error propagation. Both intra and inter predictions are utilized in popular video encoders, where an intra-coded frame only exploits the spatial redundancy within current frame, while an inter-coded frame jointly explores the spatio-temporal correlations across frames over time. 

Earlier explorations of this subject have mainly focused on designing DNN-based filters for intra-coded frames, particularly by trading network depth and parameters for better coding efficiency. For example, IFCNN~\cite{IFCNN}, and VRCNN~\cite{VRCNN} are shallow networks with $\approx$50,000 parameters, providing up to 5\% BD-Rate savings for the H.265/HEVC intra encoder. More gains can be obtained if we use a deeper and denser network~\cite{RHCNN, PCS_inception, PCS_Resnet}, {\it e.g.}, 5.7\% BD-Rate gain reported in~\cite{RHCNN} by using the model with 3,340,000 parameters, and 8.50\% BD-Rate saving obtained in~\cite{MMS-net} by using the model with 2,298,160 parameters.
The more parameters a model has, the more complex it is. Unfortunately, greater complexity limits the network's potential for practical application. Such intra-frame-based in-loop filters treat decoded frames equally, without the consideration of in-loop inter-prediction dependency. Nevertheless, aforementioned networks can be used in post-filtering out of the coding loop.

It is necessary to include temporal prediction dependency while designing the in-loop CNN-based filters for inter-frame coding. Some studies leveraged prior knowledge from the encoding process to assist the CNN training and inference. For example, Jia \emph{et al.}~\cite{STResNet} incorporated the co-located block information for in-loop filtering. Meng \emph{et al.}~\cite{MLSDRN} utilized the coding unit partition for further performance improvement. Li~\emph{et al.}~\cite{Daowen_ISCAS} input both the reconstructed frame and the difference between the reconstructed and predicted pixels to improve the coding efficiency. Applying prior knowledge in learning may improve the coding performance, but it further complicates the CNN  model by involving  additional information in the networks. On the other hand, the contribution of this prior knowledge is quite limited because such additional priors are already implicitly embedded in the reconstructed frame.

If a CNN-based in-loop filtering is applied to frame $I_0$, the impact will be gradually propagated to frame $I_1$ that has frame  $I_0$ as the reference. Subsequently, $I_1$ is the reference of $I_2$, and so on so forth\footnote{Even though more advanced inter referencing strategies can be devised, inter propagation-based behavior remains the same.}. If frame $I_1$ is filtered again by the same CNN model, an over-filtering problem will be triggered, resulting in severely degraded performance, as analyzed in~\cite{ding2019switchable}. To overcome this challenging problem, a CNN model called SimNet was built to carry the relationship between the reconstructed frame and its original frame in~\cite{ding2019cnn} to adaptively skip filtering operations in inter coding. 
SimNet reported 7.27\% and 5.57\% BD-Rate savings for intra- and inter- coding of AV1, respectively. A similar skipping strategy was suggested by Chen \etal~\cite{chen2019av1} to enable a wide activation residual network, yielding $14.42\%$ and 9.64\% BD-Rate savings for respective intra- and inter- coding on AV1 platform.     

Alternative solutions resort to the more expensive R-D optimization to avoid the over-filtering problem. For example, Yin \etal~\cite{JVET_Intel} developed three sets of CNN filters for luma and chroma components, where the R-D optimal CNN model is used and signaled in bitstream. Similar ideas are developed in~\cite{MF_inloopFilter,Content_aware_inloopfilter} as well, in which multiple CNN models are trained and the R-D optimal model is selected for inference. 

It is impractical to use deeper and denser CNN models in applications. It is also very expensive to conduct R-D optimization to choose the optimal one from a set of pre-trained models. Note that a limited number of pre-trained models are theoretically insufficient to be generalized for large-scale video samples.   
To this end, in Section~\ref{subsec:proposed_in_loop}, we introduce a guided-CNN scheme which adapts shallow CNN models according to the characteristics of input video content.

\subsection{Post Filtering}
\label{subsec:out-loop filtering}
Post filtering is generally applied to the compressed frames at the decoder side to further enhance the video quality for better QoE.

Previous in-loop filters designated for intra-coded frames can be re-used for {\it single-frame} post-filtering~\cite{dong2015compression,cavigelli2017cas,guo2016building,galteri2017deep,liu2018multi,zhang2018adaptive,VRCNN,wang2017novel,yang2017decoder,he2018enhancing}. Appropriate re-training may be applied in order to better capture the data characteristics.  
However, single-frame post-filtering may introduce quality fluctuation across frames. This may be due to the limited capacity of CNN models to deal with a great amount of video contents. Thus, {\it multi-frame} post filtering can be devised to massively exploit the correlation across successive temporal frames. By doing so, it not only greatly improves the single-frame solution, but also offers better temporal quality over time.   

Typically, a two-step strategy is applied for multi-frame post filtering. First, neighboring frames are aligned to the current frame via (pixel-level) motion estimation and compensation (MEMC). Then, all aligned frames are fed into networks for high-quality reconstruction. Thus, the accuracy of MEMC greatly affects reconstruction performance. In applications, learned optical flow, such as  FlowNet~\cite{dosovitskiy2015flownet}, FlowNet2~\cite{ilg2017flownet}, PWC-Net~\cite{sun2018pwc}, and TOFlow~\cite{xue2019video}, are widely used.

Some exploration has already been made in this arena: Bao \etal~\cite{bao2019memc} and Wang \etal~\cite{wang2019edvr} implemented a general video quality enhancement framework for denoising, deblocking, and super-resolution, where Bao \etal~\cite{bao2019memc} employed the FlowNet and Wang \etal~\cite{wang2019edvr} used pyramid, cascading, and deformable convolutions to respectively align frames temporally. Meanwhile, Yang \etal~\cite{yang2018multi} proposed a multi-frame quality enhancement framework called MFQE-1.0, in which a spatial transformer motion compensation (STMC) network is used for alignment, and a deep quality enhancement network (QE-net) is employed to improve reconstruction quality. Then, Guan \etal~\cite{guan2019mfqe} upgraded MFQE-1.0 to MFQE-2.0 by replacing QE-net using a dense CNN model, leading to better performance and less complexity. Later on, Tong \etal~\cite{tong2019learning} suggested using FlowNet2 in MFQE-1.0 for temporal frame alignment (instead of default STMC), yielding 0.23 dB PSNR gain over the original MFQE-1.0. Similarly, FlowNet2 is also used in~\cite{lu2019learned} for improved efficiency.

All of these studies suggested the importance of temporal alignment in post filtering. Thus, in the subsequent case study (see Section~\ref{subsec:proposed_out_loop}), we first examine the efficiency of alignment, and then further discuss the contributions from respective intra-coded and inter-coded frames for the quality enhancement of final reconstruction. This will help audiences gain a deeper understanding of similar post filtering techniques.

	\section{Case Study for Pre-processing:\\ Switchable Texture-based Video Coding}
\label{sec:proposed_pre_processing}

This section presents a switchable texture-based video pre-processing that leverages DNN-based semantic understanding for subsequent coding improvement. In short, we exploit DNNs to accurately segment ``perceptually InSIGnifcant'' (pInSIG) texture  areas to produce a corresponding pInSIG mask. In many instances, this mask drives the encoder to perform separately for pInSIG textures that are typically inferred without additional residuals, and ``perceptually SIGnificant'' (pSIG) areas elsewhere using traditional hybrid coding method. This approach is implemented on top of the AV1 codec \cite{av1-joshi2017,av1-liu2017,av1-chen2018} by enabling the GoP-level switchable mechanism, This yields noticeable bit rate savings for both standard test sequences and additional challenging sequences from YouTube UGC dataset \cite{wang2019ugc}, under similar perceptual quality. The method we propose is a pioneering work that integrates learning-based texture analysis and reconstruction approaches with modern video codec to enhance video compression performance.

\begin{figure}[t]
	\centering
	\centerline{\includegraphics[width=\linewidth]{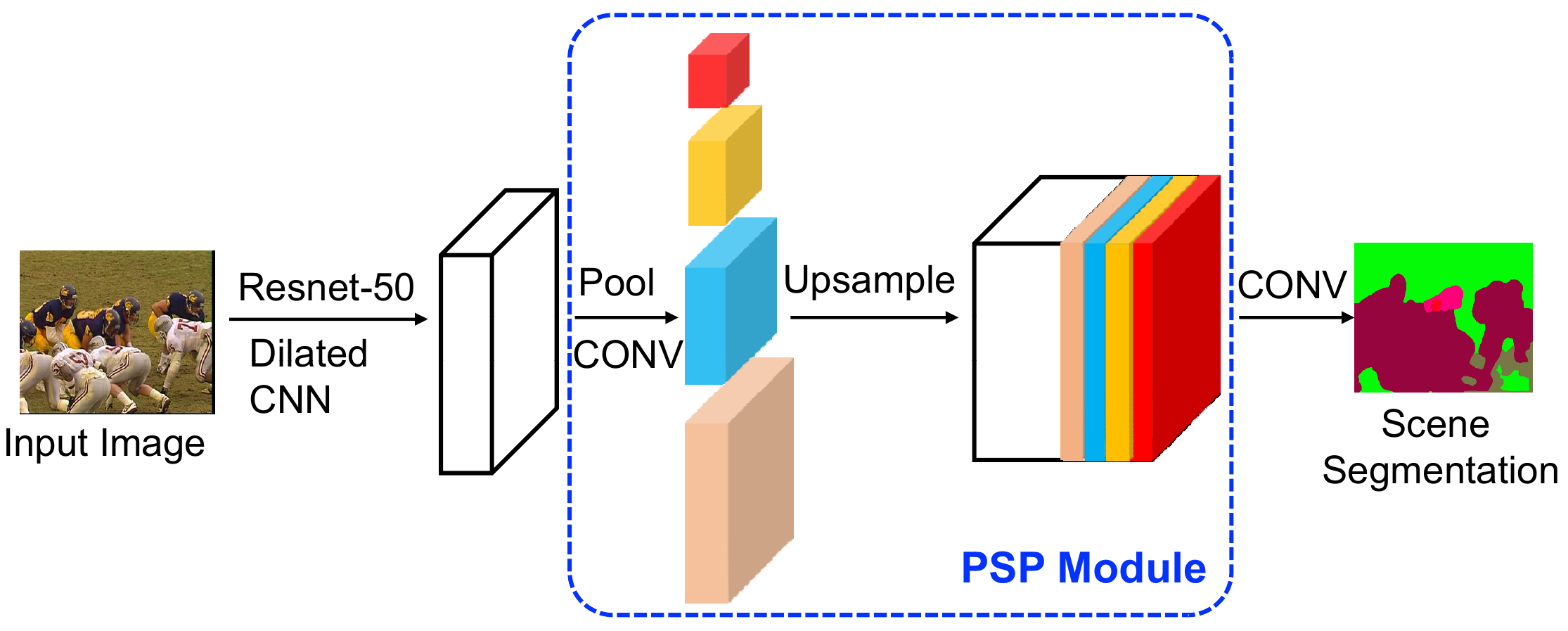}}
	\caption{{\bf Texture Analyzer.} {Proposed semantic segmentation network using PSPNet~\cite{zhao2017pyramid} and ResNet-50~\cite{he2016deep}.}}
	\label{fig:sec5_texture_preprocessing_network}
\end{figure}

\subsection{Texture Analysis}
\label{subsec:texture_analysis}
Our previous attempt~\cite{bosch-jstsp2011} yielded encouraging bit rate savings without decreasing visual quality. This was accomplished by perceptually differentiating pInSIG textures and other areas to be encoded in a hybrid coding framework. However, the corresponding texture masks were derived using traditional methods, at the coding block level. 
{On the other hand, building upon advancements created by DNNs and large-scale labeled datasets ({\it e.g.}, ImageNet~\cite{russakovsky2015imagenet}, COCO~\cite{lin2014microsoft}, and ADE20K~\cite{zhou2019semantic}), learning-based semantic scene segmentation algorithms~\cite{long2015fully, zhao2017pyramid, zhou2019semantic}  have been tremendously improved to generate accurate pixel-level texture masks. 

In this work, we first rely on the powerful ResNet50~\cite{he2016deep} with dilated convolutions~\cite{chen2014semantic,yu2015multi} to extract feature maps that effectively embed the content semantics. We then introduce the pyramid pooling module from PSPNet~\cite{zhao2017pyramid} to produce a pixel-level semantic segmentation map shown in Fig.~\ref{fig:sec5_texture_preprocessing_network}.  Our implementation starts with a pre-trained PSPNet model generated using the MIT SceneParse150~\cite{zhou2017scene} as a scene parsing benchmark. We then retrained the model on a subset of a densely annotated dataset ADE20K~\cite{zhou2019semantic}. In the end, the model offers a pixel segmentation accuracy of 80.23\%. 

It is worthwhile to note that such pixel-level segmentation may result in the creation of a number of semantic classes. Nevertheless, this study suggests grouping similar texture classes commonly found in nature scenes together into four major categories, {\it e.g.}, ``earth and grass'', ``water, sea and river'', ``mountain and hill'', and ``tree''. Each texture category would have an individual segmentation mask to guide the compression performed by the succeeding video encoder.

\subsection{Switchable Texture-Based Video Coding}
\label{subsec:texture_mode}
Texture masks are generally used to identify texture blocks, and to perform the encoding of texture blocks and non-texture blocks separately, as illustrated in Fig.~\ref{fig:texture_mode}. In this case study, the AV1 reference software platform is selected to exemplify the efficiency of our proposal.

{\bf Texture Blocks.} Texture and non-texture blocks are identified by overlaying the segmentation mask from the texture analyzer on its corresponding frame. These frame-aligned texture masks produce pixel-level accuracy, which is capable of supporting arbitrary texture shapes. However, in order to support the block processing commonly adopted by video encoders, we propose refining original pixel-level masks to their block-based representations.  The minimum size of a texture block is 16$\times$16. 
In order to avoid boundary artifacts and maintain temporal consistency, we implemented a conservative two-step strategy to determine the texture block. First, the block itself must be fully contained in the texture region marked using the pixel-level mask. Then, its warped representation to temporal references ({\it e.g.}, the preceding and succeeding frames in the encoding order) have to be inside the  masked texture area of corresponding reference frames as well. Finally, these texture blocks are encoded using the {\it texture mode}, and non-texture blocks are encoded as usual using the hybrid coding structure.

{\bf Texture Mode.} 
A texture mode coded block is inferred by its temporal reference using the global motion parameters without incurring any motion compensation residuals. In contrast, non-texture blocks are compressed using a hybrid ``prediction+residual'' scheme.
For each current frame and any one of its reference frames, AV1 syntax specifies only one set of global motion parameters at the frame header.
Therefore, to comply with the AV1 syntax, our implementation only considers one texture class for each frame. This guarantees the general compatibility of our solution to existing AV1 decoders. 
We further modified the AV1 global motion tool to estimate the motion parameters based on the texture regions of the current frame and its reference frame. We used the same feature extraction and model fitting approach as in the global motion coding tool in order to provide a more accurate motion model for the texture regions. This was done to prevent visual artifacts on the block edges between the texture and non-texture blocks in the reconstructed video.
Although we have demonstrated our algorithms using the AV1 standard, we expect that the same methodology can be applied to other standards. For instance, when using the H.265/HEVC standard,  we can leverage the SKIP mode syntax to signal the texture mode instead of utilizing the global motion parameters.
 
Previous discussions have suggested that the texture mode is enabled along with inter prediction. Our extensive studies have also demonstrated that it is better to activate the texture mode in frames where bi-directional predictions are allowed ({\it e.g.}, B-frames), for the optimal trade-off between bit rate saving and perceived quality. As will be shown in following performance comparisons, we use a 8-frame GoP (or Golden-Frame (GF) group defined in AV1) to exemplify the texture modes in every other frame, by which the compound prediction from bi-directional references can be facilitated for prediction warping. Such bi-directional prediction could also alleviate possible temporal quality flickering.

{\bf Switchable Optimization.} 
In our previous work~\cite{Di2019}, the texture mode was enabled for every B frame, demonstrating significant bit rate reduction at the same level of perceptual sensation in most standard test videos, in comparison to the AV1 anchor. However, some videos did cause the model to perform more poorly. One reason for this effect is that higher QP settings typically incur more all-zero residual blocks. Alternatively, texture mode is also content-dependent: a relatively small number of texture blocks may be present for some videos. Both scenarios limit the bit rate savings, and an overhead of extra bits is mandatory for global motion signaling, if texture mode is enabled.

To address these problems, we introduce a switchable scheme to determine whether texture mode could be potentially enabled for a GoP or a GF group. The criteria for switching are based on the texture region percentage that is calculated as the average ratio of texture blocks in B-frames, and on the potential bit rate savings with or without texture mode. Figure~\ref{fig:switchable_scheme_of_texture_mode} illustrates the switchable texture mode decision. Currently, we use bit rate saving as a criterion for switch decisions when  the texture mode is enabled. This assumes perceptual sensation will remain nearly the same, since these texture blocks are perceptually insignificant.  

\begin{figure}[t]
     \centering
     \subfloat[]{\includegraphics[width=\linewidth]{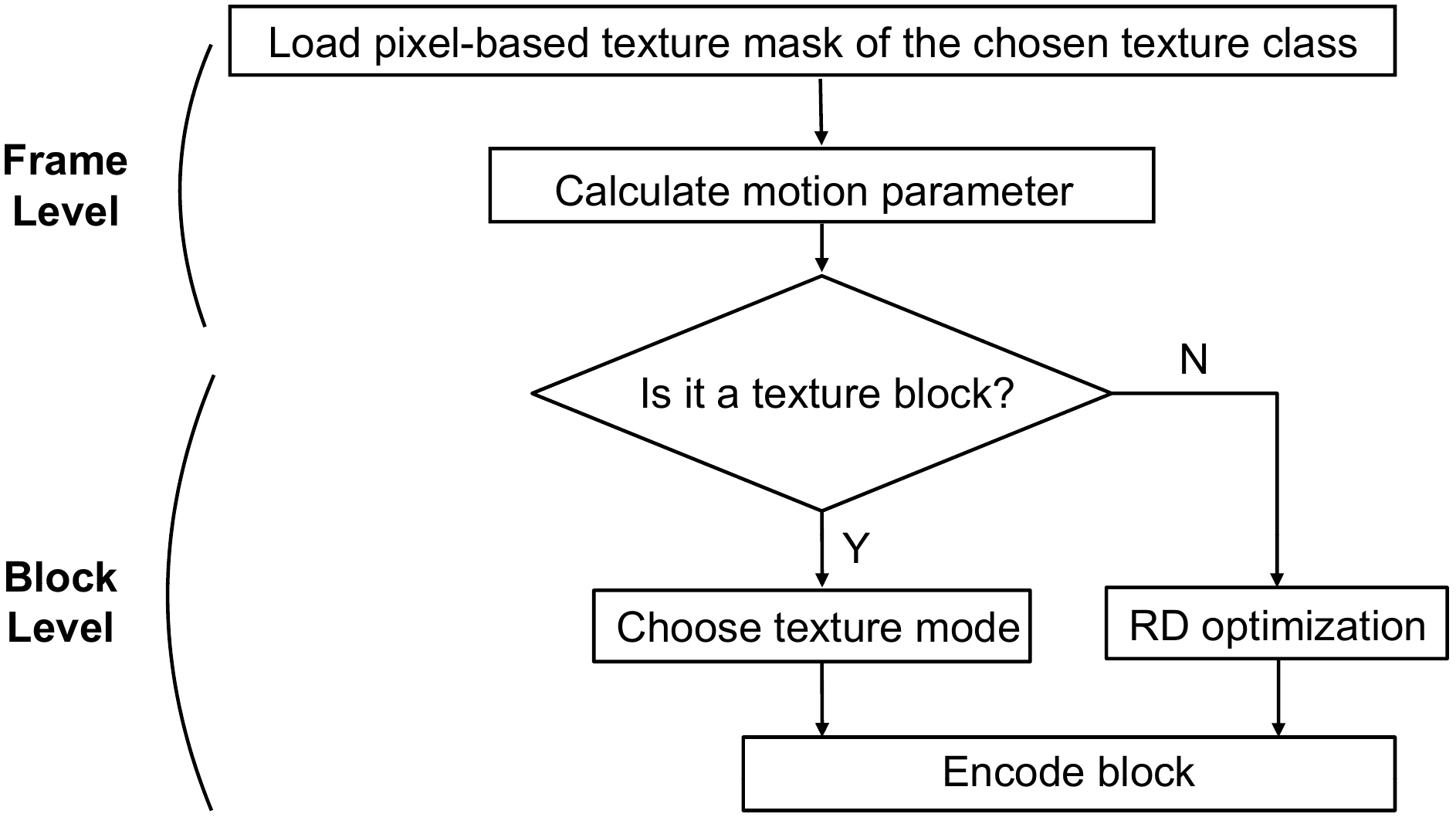} \label{fig:texture_mode}}\\
     \subfloat[]{\includegraphics[width=\linewidth]{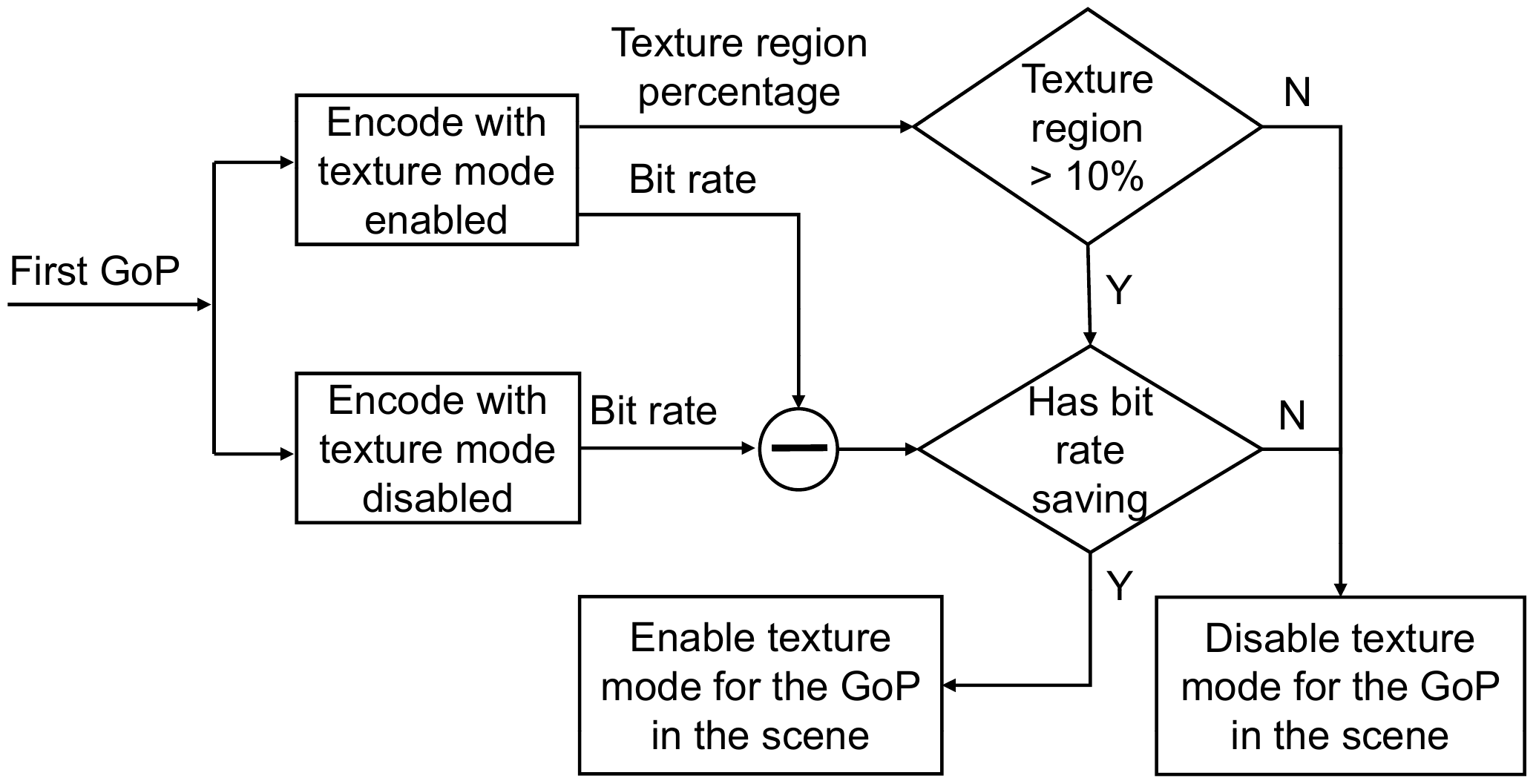} \label{fig:switchable_scheme_of_texture_mode}}
     \caption{{\bf Texture mode and switchable control scheme.} (a) Texture mode encoder implementation. (b) Switchable texture mode decision. }
\end{figure}

\subsection{Experimental Results}
\label{subsec:results}
We selected sequences with texture regions from standard test sequences and the more challenging YouTube UGC data set\footnote{\url{https://media.withyoutube.com/}} \cite{wang2019ugc}. YouTube UGC dataset is a sample selected from thousands of User Generated Content (UGC) videos uploaded to YouTube. The names of the UGC videos follow the format of Category\_Resolution\_UniqueID. We calculate the bit rate savings at different QP values for 150 frames of the test sequences. 
In our experiments, we used the following parameters for the AV1 codec\footnote{AV1 codec change-Id: Ibed6015aa7cce12fcc6f314ffde76624df4ad2a1} as the baseline: 8-frame GoP or GF group using random access configuration; 30 FPS; constant quality rate control policy; multi-layer coding structure for all GF groups; maximum intra frame interval at 150.}
We evaluate the performance of our proposed method in terms of bit rate savings and perceived quality.

\subsubsection{Coding Performance}
To evaluate the performance of the proposed switchable texture mode method, bit rate savings at four quantization levels (QP = 16, 24, 32, 40) are calculated for each test sequence in comparison to the AV1 baseline. 

\begin{table*}[!htbp]
	\caption{Bit rate saving (\%) comparison between handcraft feature (FM)~\cite{bosch2007}, block-level DNN (BM)~\cite{fu2018} and pixel-level DNN (PM)~\cite{Di2019} texture analysis against the AV1 baseline for selected standard test sequences using \textit{tex-allfg} method.}
	\centering
	\label{tab:texture_compare}
	{
		\begin{tabular}{c|d{3.2} d{3.2} d{3.2} | d{3.2} d{3.2} d{3.2} | d{3.2} d{3.2} d{3.2} | d{3.2} d{3.2} d{3.2}}
			\hline
			\multirow{2}{*}{{Video Sequence}} & \multicolumn{3}{c|}{{QP=16} (\%)} & \multicolumn{3}{c|}			{{QP=24} (\%)} & \multicolumn{3}{c|}{{QP=32} (\%)} & \multicolumn{3}{c}{{QP=40} (\%)} \\ 
			& \multicolumn{1}{c}{FM} & 
			\multicolumn{1}{c}{BM} & 
			\multicolumn{1}{c|}{PM} & 
			\multicolumn{1}{c}{FM} & 
			\multicolumn{1}{c}{BM} & 
			\multicolumn{1}{c|}{PM} & 
			\multicolumn{1}{c}{FM} & 
			\multicolumn{1}{c}{BM} & 
			\multicolumn{1}{c|}{PM} & 
			\multicolumn{1}{c}{FM} & 
			\multicolumn{1}{c}{BM} & 
			\multicolumn{1}{c}{PM} \\ \hline
			Coastguard & -0.17 & 7.80   & 9.14 & -0.36 & 6.99 & 8.01 & -0.43 & 4.70  & 5.72 & -0.62 & 1.90 & 		2.13 \\ \hline
			Flower & 7.42 & 10.55 & 13.00   & 5.42 & 8.66 & 10.78 & 2.51 & 5.96 & 4.95 & 0.19 & 3.38 & 			1.20 \\ \hline
            Waterfall & 3.65 & 4.63  & 13.11  & 1.58 & 3.96 & 7.21  & -0.14 & -0.33  & 1.30  & -3.00 & 			-3.74  & -3.48 \\ \hline
            Netflix\_aerial & 1.15 & 8.59 & 9.15   & -0.26 & 2.15 & 5.59  & -1.32 & -0.68   & 1.05  & -2.10 	& -4.59  & -4.01 \\ \hline
			Intotree & 0.88 & 5.32  & 9.71   & 0.15 & 4.32  & 9.42   & -0.14 & 1.99  & 8.46 & -0.26 & -2.83    & 4.92 \\ \hline
		\end{tabular}
	}
\end{table*}

\textbf{Texture Analysis.}
We compare two DNN-based texture analysis methods~\cite{fu2018,Di2019} with a handcrafted feature-based approach~\cite{bosch2007} for selected standard test sequences.  Results are shown in Table~\ref{tab:texture_compare}. A positive bit rate saving (\%) indicates a reduction compared with the AV1 baseline. Compared to the feature based approach, DNN-based methods show improved performance in terms of bit rate saving. The feature based approach relies on color and edge information to generate the texture mask and is less accurate and consistent both spatially and temporally. Therefore, the number of blocks that are reconstructed using texture mode is usually much smaller than that of DNN-based methods. Note that the parameters used in feature based approach require manually tuning for each video to optimize the texture analysis output. The pixel-level segmentation~\cite{Di2019} shows further advantages compared with block-level method~\cite{fu2018}, since the CNN model does not require block size to be fixed.

\begin{table*}[!htbp]
	\renewcommand{\arraystretch}{1.06} 
	\caption{Bit rate saving (\%) comparison for \textit{tex-allgf} and \textit{tex-switch} methods against the AV1 baseline.}
	\centering
	\label{tab:ugc_datarate}
	{
		\begin{tabular}{c c|d{3.2} d{3.2} | d{3.2} d{3.2} | d{3.2} d{3.2} | d{3.2} d{3.2}}
			\hline
			\multirow{2}{*}{{Resolution}} & \multirow{2}{*}{{Video Sequence}} & \multicolumn{2}{c|}{{QP=16} (\%)} & \multicolumn{2}{c|}{{QP=24} (\%)} & \multicolumn{2}{c|}{{QP=32} (\%)}      & \multicolumn{2}{c}{{QP=40} (\%)}      \\ 
			& & \multicolumn{1}{c}{\textit{tex-allgf}} & \multicolumn{1}{c|}{\textit{tex-switch}} & \multicolumn{1}{c}{\textit{tex-allgf}} & \multicolumn{1}{c|}{\textit{tex-switch}} &
			\multicolumn{1}{c}{\textit{tex-allgf}} & \multicolumn{1}{c|}{\textit{tex-switch}} & \multicolumn{1}{c}{\textit{tex-allgf}} & \multicolumn{1}{c}{\textit{tex-switch}} \\ \hline
			\multirow{5}{*}{CIF} & Bridgeclose  & 15.78  & 15.78 & 10.87 & 10.87 & 4.21  & 4.21  & 2.77  & 2.77 \\ 
			& Bridgefar  & 10.68  & 10.68 & 8.56  & 8.56  & 6.34  & 6.34  & 6.01  & 6.01 \\ 
			& Coastguard  & 9.14   & 9.14  & 8.01  & 8.01  & 5.72  & 5.72  & 2.13  & 2.13 \\ 
			& Flower  & 13.00  & 13.00 & 10.78 & 10.78 & 4.95  & 4.95  & 1.20  & 1.20 \\ 
			& Waterfall  & 13.11  & 13.11 & 7.21  & 7.21  & 1.30  & 1.30  & \cellcolor{newgreen}-3.48  & \cellcolor{newgreen}0.00 \\  \hline
			512$\times$270  & Netflix\_ariel  & 9.15 & 9.15 & 5.59  & 5.59  & 1.05  & 1.05  & \cellcolor{newgreen}-4.01  & \cellcolor{newgreen}0.00\\  \hline
			\multirow{5}{*}{360P}  & NewsClip\_360P-1e1c  & 10.77  & 10.77  & 9.27  & 9.27  & 5.23  & 5.23  & 1.54  & 1.54 \\ 
			& NewsClip\_360P-22ce  & 17.37  & 17.37  & 15.79  & 15.79  & 16.37  & 16.37  & 17.98  & 17.98 \\ 
			& TelevisionClip\_360P-3b9a  & 1.45  & 1.45  & 0.48  & 0.48  & \cellcolor{newgreen}-1.09  & \cellcolor{newgreen}0.00  & \cellcolor{newgreen}-3.26  & \cellcolor{newgreen}0.00 \\ 
			& TelevisionClip\_360P-74dd  & 1.66  & 1.66  & 1.17  & 1.17  & 0.36  & 0.36  & \cellcolor{newgreen}-0.37  & \cellcolor{newgreen}0.00 \\  \hline
			\multirow{6}{*}{480P}  & HowTo\_480P-04f1  & 3.81  & 3.81  & 2.57  & 2.57  & 0.93  & 0.93  & \cellcolor{newgreen}0.06  &\cellcolor{newgreen}0.36 \\ 
			& HowTo\_480P-4c99  & 2.36  & 2.36  & 1.67  & 1.67  &\cellcolor{newred} 0.37  & \cellcolor{newred}0.00  & \cellcolor{newgreen}-1.16  & \cellcolor{newgreen}0.00 \\ 
			& MusicVideo\_480P-1eee  & 3.31  & 3.31  & 3.29  & 3.29  & 2.53  & 2.53  & -0.30  & -0.30 \\ 
			& NewsClip\_480P-15fa  & 6.31  & 6.31  &\cellcolor{newred} 6.05  &\cellcolor{newred} 5.79  &\cellcolor{newred} 0.53  & \cellcolor{newred}0.11  & \cellcolor{newgreen}-0.79  & \cellcolor{newgreen}0.03 \\ 
			& NewsClip\_480P-7a0d  & 11.54  & 11.54  & 10.03  & 10.03  & 1.53  & 1.53  & \cellcolor{newred}0.08  &\cellcolor{newred} 0.00 \\ 
			& TelevisionClip\_480P-19d3  & 3.13  & 3.13  & 2.86  & 2.86  & 1.66  & 1.66  & \cellcolor{newred}0.58  &\cellcolor{newred} 0.00 \\  \hline
			\multirow{3}{*}{720P}  & HowTo\_720P-0b01  & 12.72  & 12.72  & 11.84  & 11.84  & 9.31  & 9.31  & 6.35  & 6.35 \\ 
			& MusicVideo\_720P-3698  & 1.76  & 1.76  & 1.07  & 1.07  & 0.30  & 0.30  & \cellcolor{newgreen}-0.17  & \cellcolor{newgreen}0.00 \\ 
			& MusicVideo\_720P-4ad2  & 6.93  & 6.93  & 3.81  & 3.81  & 1.87  & 1.87  &\cellcolor{newred}0.60  & \cellcolor{newred}0.11 \\ \hline
			\multirow{3}{*}{1080P}  & HowTo\_1080P-4d7b  & 7.31  & 7.31  & 6.07  & 6.07  & 3.21  & 3.21  & 0.72  & 0.72 \\ 
			& MusicVideo\_1080P-55af  & 3.88  & 3.88  & 1.78  & 1.78  &\cellcolor{newgreen} 0.31  & \cellcolor{newgreen}0.33  &\cellcolor{newgreen} -0.99  &\cellcolor{newgreen} -0.68 \\ 
			& intotree  & 9.71  & 9.71  & 9.42  & 9.42  & 8.46  & 8.46  & 4.92  & 4.92 \\ \hline
			& {Average} & 7.96	& 7.96	& \cellcolor{newred}6.28	& \cellcolor{newred}6.27	& \cellcolor{newgreen}3.38	& \cellcolor{newgreen}3.40	& \cellcolor{newgreen}1.45	& \cellcolor{newgreen}2.05 \\
			\hline
		\end{tabular}
	}
\end{table*}

\textbf{Switchable Scheme.}
We also compare the proposed method, a.k.a., \textit{tex-switch}, with our previous work in \cite{Di2019}, a.k.a., \textit{tex-allgf}, which enables texture mode for all frames in a GF group. All three methods use the same encoder setting for fair comparison.
Bit rate saving results for various videos at different resolutions against the AV1 baseline are shown in Table~\ref{tab:ugc_datarate}. A positive bit rate saving (\%) indicates a reduction compared with the AV1 baseline.

In general, compared to the AV1 baseline, the coding performance of \textit{tex-allgf} shows significant bit rate savings at lower QPs. However, as QP increases, the savings are diminished. In some cases, \textit{tex-allgf} exhibits poorer coding performance than the AV1 baseline at a high QP ({\it e.g.}, negative numbers at QP 40). At a high QP, most blocks have zero residual due to heavy quantization, leading to very limited margins for bit rate savings using texture mode. In addition, few extra bits are required for the signalling of global motion of texture mode coded blocks. The bit savings gained through residual skipping in texture mode still cannot compensate for the bits used as overhead for the side information.

Furthermore, the proposed \textit{tex-switch} method retains the greatest bit rate savings offered by \textit{tex-allgf}, and resolves the loss at higher QP settings. As shown in Table~\ref{tab:ugc_datarate}, negative numbers are mostly removed (highlighted in green) by the introduction of a GoP-level switchable texture mode.
In some cases where \textit{tex-switch} has zero bit rate savings compared to the AV1 baseline, the texture mode is completely disabled for all the GF groups, whereas \textit{tex-allgf} has loss. In a few cases, however, \textit{tex-switch} has less bit rate saving than \textit{tex-allgf} (highlighted in red). This is because the bit rate saving performance of the first GF group in the scene fails to accurately represent the whole scene in some of the UGC sequences with short scene cuts. A possible solution is to identify additional GF groups that show potential bit rate savings and enable texture mode for these GF groups. 

\subsubsection{Subjective Evaluation}
\label{ssec:subjective}

\begin{figure}[t]
	\centering
	\centerline{\includegraphics[width=\linewidth]{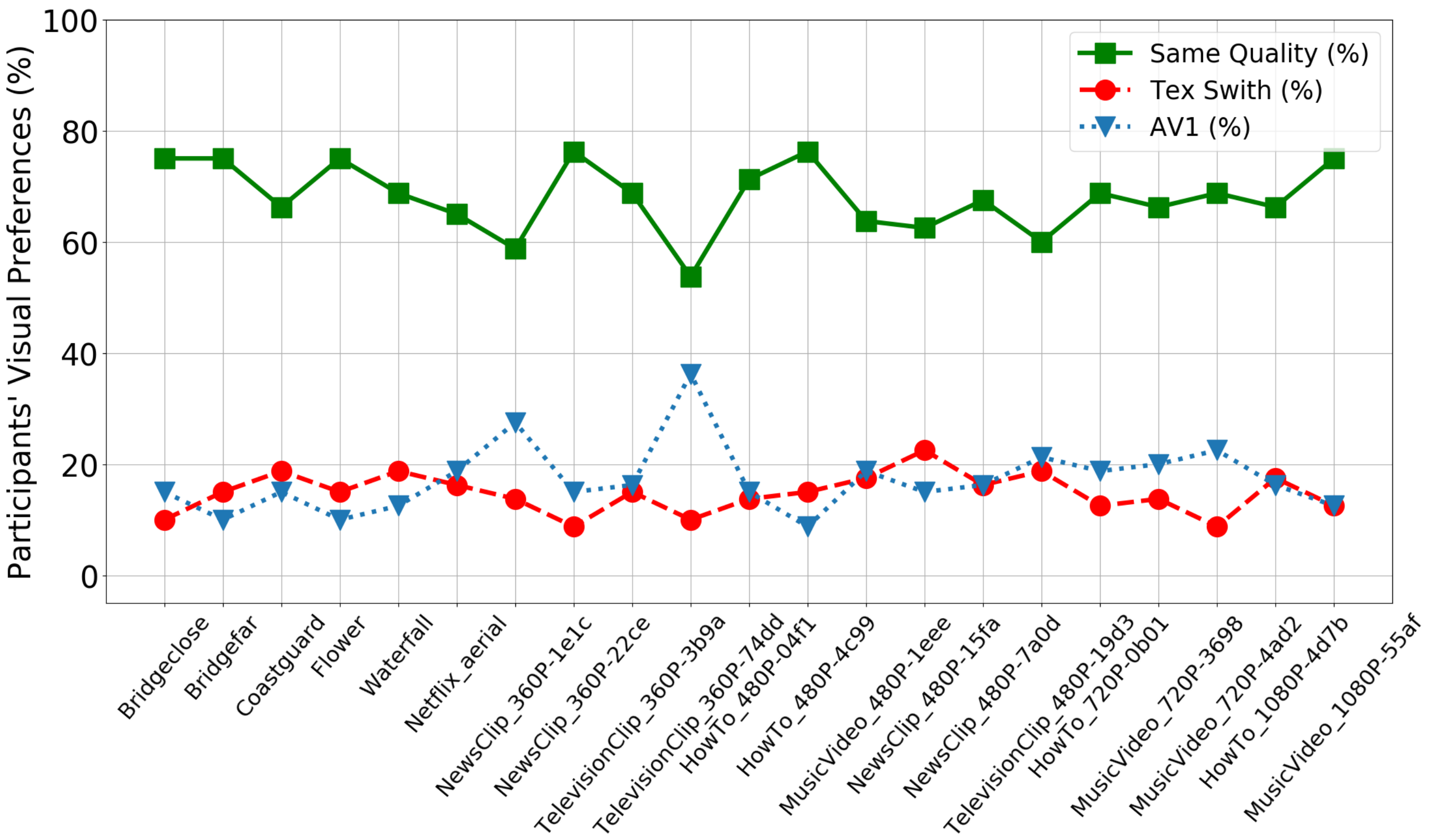}}
	\caption{{\bf Subjective evaluation of visual preference.} Results show average subjective preference (\%) for QP = 16, 24, 32, 40 compared between AV1 baseline and proposed switchable texture mode.}
	\label{fig:subjective}
\end{figure}

Although significant bit rate savings have been achieved compared to the AV1 baseline, it is acknowledged that identical QP values do not necessarily imply the same video quality. We have performed a subjective visual quality study with 20 participants. Reconstructed videos produced by the proposed method (\textit{tex-switch}) and the baseline AV1 codec at QP = 16, 24, 32 and 40 are arranged randomly and assessed by the participants using a double stimulus continuous quality scale (DSCQS) method \cite{itu2019}. Subjects have been asked to choose among three options: the first video has better visual quality, the second video has better visual quality, or there is no difference between two versions.

The result of this study is summarized in Figure \ref{fig:subjective}. The ``Same Quality" indicates the percentage of participants that cannot tell the difference between the reconstructed videos by the AV1 baseline codec and the proposed method \textit{tex-switch} (69.03\% on average). The term ``tex-switch'' indicates the percentage of participants that prefer the reconstructions by the proposed method \textit{tex-switch} (14.32\% on average); and the ``AV1'' indicates the percentage of participants who think the visual quality of the reconstructed videos using the AV1 baseline is better (16.65\% on average). 

We observe that the results are sequence dependent and that spatial and temporal artifacts can appear in the reconstructed video. 
The main artifacts come from the inaccurate pixel-based texture mask. For example, in some frames of \textit{TelevisionClip\_360P-74dd} sequence, the texture masks include parts of the moving objects in the foreground, which are reconstructed using texture mode. Since the motion of the moving objects is different from the motion of the texture area, there are noticeable artifacts around those parts of the frame. To further improve the accuracy of region analysis using DNN-based pre-processing, we plan to incorporate an in-loop perceptual visual quality metric for optimization during the texture analysis and reconstruction.  

\subsection{Discussion And Future Direction}

We proposed a DNN based texture analysis/synthesis coding tool for AV1 codec. Experimental results show that our proposed method can achieve noticeable bit rate reduction with satisfying visual quality for both standard test sets and user generated content, which is verified by a subjective study. 
We envision that video coding driven by semantic understanding will continue to improve in terms of both quality and bit rate, especially by leveraging advances of deep learning methods. However, there remain several open challenges that require further investigation.

Accuracy of region analysis is one of the major challenges for integrating semantic understanding into video coding. However, recent advances in scene understanding have significantly improved the performance of region analysis. 
Visual artifacts are still noticeable when a non-texture region is incorrectly included in the texture mask, particularly if the analysis/synthesis coding system is open loop. One potential solution is to incorporate some perceptual visual quality measures in-loop during the texture region reconstruction.

Video segmentation benchmark datasets are important for developing machine learning methods for video based semantic understanding. Existing segmentation datasets are either based on images with texture~\cite{HaiMik08b}, or contain general video objects only~\cite{Ning2018dataset,perazzi2016benchmark}, or focus on visual quality but lack segmentation ground truth. 


\section{Case Study for Coding:\\ End-to-End Neural Video Coding (E2E-NVC)}
\label{sec:proposed_end_to_end_coding}

This section presents a framework for end-to-end neural video coding. We include a discussion of its key components, as well as its overall efficiency.  {Our proposed method is extended from  our  pioneering  work in~\cite{chen2017deepcoder} but with significant performance improvements by allowing fully end-to-end learning-based spatio-temporal feature representation.
More details can be found in~\cite{chen2019neural,liu2020neural, liu2020learned}.

\begin{figure}[t]
     \centering
     \subfloat[]{\includegraphics[width=\linewidth]{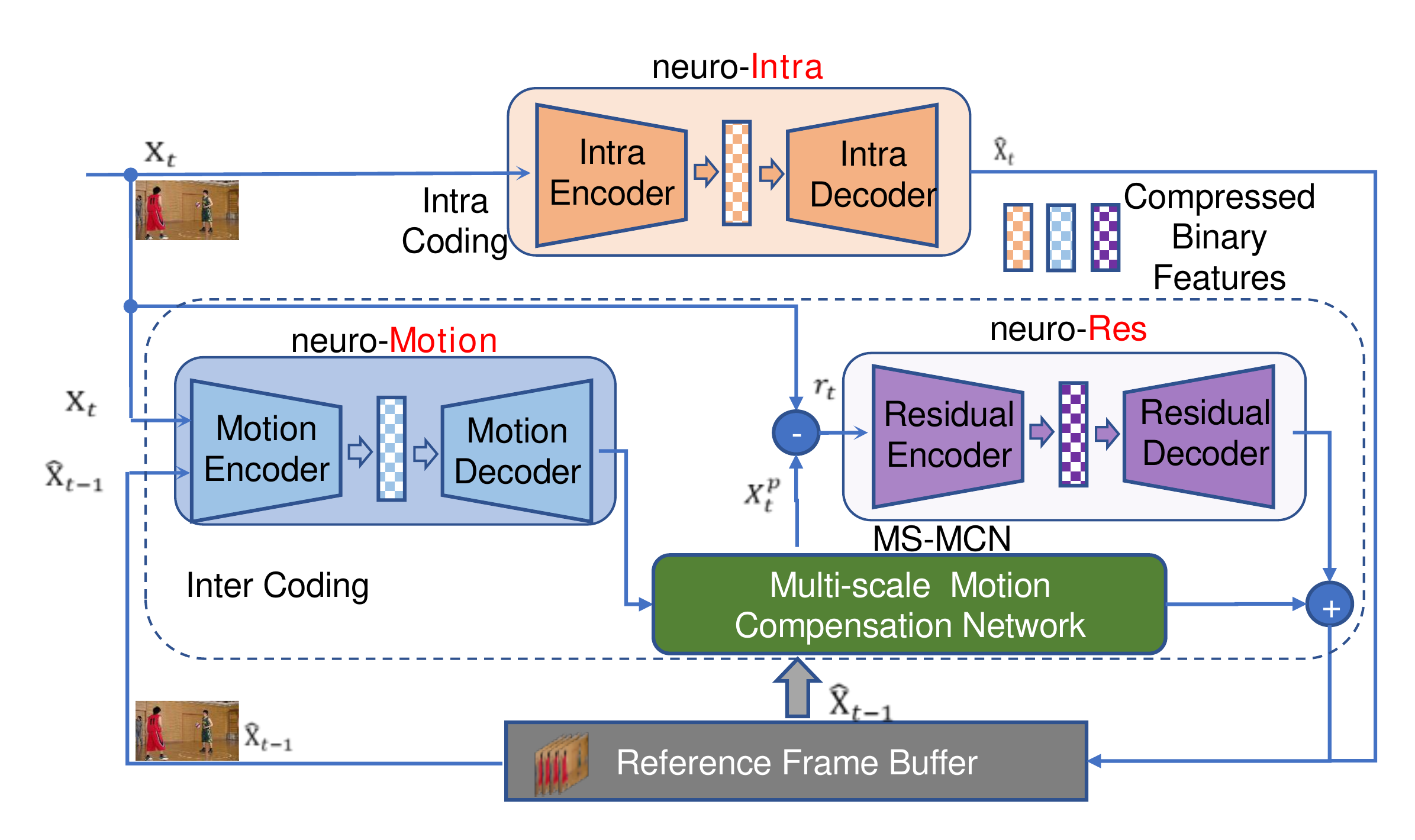} \label{fig:nvc_architecture}}\\
     \subfloat[]{\includegraphics[width=\linewidth]{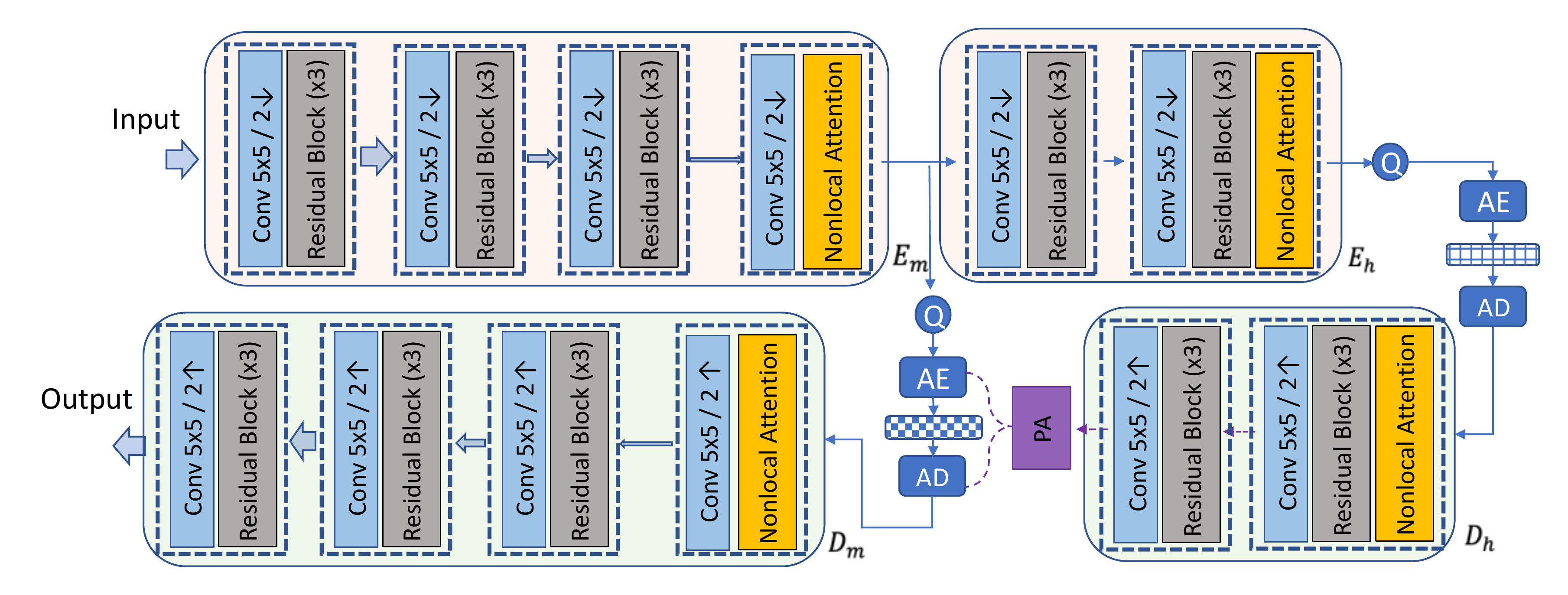} \label{fig:bio_video_coder}}
     \caption{{\bf End-to-End Neural Video Coding (E2E-NVC).} This E2E-NVC in (a) consists of modularized intra and inter coding, where inter coding utilizes respective motion and residual coding. Each component is well exploited using a stacked CNNs-based VAE for efficient representations of intra pixels, displaced inter residuals, and inter motions. All modularized components are inter-connected and optimized in an end-to-end manner. (b) General VAE model applies stacked convolutions ({\it e.g.}, 5$\times$5) with main encoder-decoder (${\bf E}_m$, ${\bf D}_m$) and hyper encoder-decoder pairs (${\bf E}_h$, ${\bf D}_h$), where main encoder ${\bf E}_m$ includes four major convolutional layers ({\it e.g.}, convolutional downsampling and three residual blocks ($\times$3) robust feature processing~\cite{he2016deep}). Hyper decoder  ${\bf D}_h$ mirrors the steps in hyper encoder ${\bf E}_h$ for hyper prior information generation. Prior aggregation (PA) engine collects the information from hyper prior, autoregressive spatial neighbors, as well as temporal correspondences (if applicable)  for main decoder ${\bf D}_m$ to reconstruct input scene. Non-local attention is adopted to simulate the saliency masking at bottlenecks, and rectified linear unit (ReLU) is implicitly embedded with convolutions for enabling the nonlinearity. ``Q'' is for quantization, AE and AD for respective arithmetic encoding and decoding. 2$\downarrow$ and 2$\uparrow$ are downsampling and upsampling at a factor of 2 for both horizontal and vertical dimensions. }
\end{figure}

\subsection{Framework}
As with all modern video encoders, the proposed E2E-NVC compresses the first frame in each group of pictures as an intra-frame using a VAE based compression engine (neuro-Intra).  It codes the remaining frames in each group using motion compensated prediction. As shown in Fig.~\ref{fig:nvc_architecture}, the proposed E2E-NVC uses the VAE compressor (neuro-Motion) to generate the multiscale motion field between the current frame and the reference frame. Then, a multiscale motion compensation network (MS-MCN) takes multiscale compressed flows, warps the multiscale features of the reference frame, and combines these warped features to generate the predicted frame.  The prediction residual is then coded using another VAE-based compressor (neuro-Res).

A low-delay E2E-NVC based video encoder is specifically illustrated in this work. Given a group of pictures (GOP) $\mathbb{X}$ = \{${\bf X_1},{\bf X_2},...,{\bf X_t}$\}, we first encode ${\bf X_1}$ using the neuro-Intra module and have its reconstructed frame $\hat{\bf X}_1$. The following frame ${\bf X}_2$ is encoded predictively, using neuro-Motion, MS-MCN, and neuro-Res together, as shown in Fig.~\ref{fig:nvc_architecture}.  
Note that MS-MCN takes the multiscale optical flows $\left\{\vec{f}^1_d,\vec{f}^2_d,...,\vec{f}^s_d\right\}$ derived by the pyramid decoder in neuro-Motion, and then uses them to generate the predicted frame $\hat{\bf X}^p_2$ by multiscale  motion compensation. Displaced inter-residual $ {\bf r}_2 = {{\bf X}_2} - {\hat{\bf X}^p_2}$ is then compressed in neuro-Res, yielding the reconstruction $\hat{\bf r}_2$. The final reconstruction $\hat{\bf X}_2$ is given by
${\hat{\bf X}_2} = {\hat{\bf X}^p_2} + {\hat{\bf r}_2}$. All of the remaining P-frames in the group of pictures are then encoded using the same procedure. 

Fig.~\ref{fig:bio_video_coder} illustrates the general architecture of the VAE model. The VAE model includes a main encoder-decoder pair that is used for latent feature analysis and synthesis, as well as a hyper encoder-decoder for hyper prior generation. The main encoder ${\bf E}_m$ uses four stacked CNN layers. Each convolutional layer employs stride convolutions to achieve downsampling (at a factor of 2 in this example) and cascaded convolutions for efficient feature extraction (here, we use three ResNet-based residual blocks~\cite{he2016deep})\footnote{We choose to apply cascaded ResNets for stacked CNNs because they are highly  efficient and reliable. Other efficient CNN architectures could also be applied.}. We use two-layer hyper encoder  ${\bf E}_h$ to further generate the subsequent hyper priors as side information, which is used in the entropy coding of the latent features.

We apply stacked convolutional layers with a limited (3$\times$3) receptive field to capture the spatial locality. These convolutional layers are stacked in order to simulate layer-wise feature extraction. These same ideas are used in many relevant studies~\cite{balle2018variational,minnen2018joint}. We utilize the simplest ReLU as the nonlinear activation function(although other nonlinear activation functions such as the Generalized Divisive Normalization could be used as well)  in~\cite{balle2016end}.

The human visual system operates in two stages: First, the observer scans an entire scene to gain a complete understanding of everything within the field of vision. Second, the observer focuses their attention on specific salient regions. During image and video compression, this mechanism of visual attention can be used to ensure that bit resources are allocated where they are most needed ({\it e.g.}, via unequal feature quantization)~\cite{li2018learning,mentzer2018conditional}. This allows resources to be assigned such that salient areas are more accurately reconstructed, while resources are conserved in the reconstruction of less-salient areas. To more accurately discern salient from non-salient areas, we adopt the non-local attention module (NLAM) at the bottleneck layers of both the main encoder and hyper encoder, prior to quantization, in order to include both global and local information. 

To enable more accurate conditional probability density modeling for entropy coding of the latent features, we introduce the {\it Prior Aggregation} (PA) engine which fuses the inputs from the hyper priors, spatial neighbors, and temporal context (if applicable)\footnote{Intra and residual coding only use joint spatial and hyper priors without temporal inference.}. Information theory suggests that more accurate context modeling requires fewer resources ({\it e.g.}, bits) to represent information~\cite{cover2012elements}. For the sake of simplicity, we assume the latent features ({\it e.g.}, motion, image pixel, residual) are following the Gaussian distribution as in~\cite{minnen2018joint,hu2020coarse}. We use the PA engine to derive the mean and standard deviation of the distribution for each feature.

\begin{figure}[t]
	\centering
	\includegraphics[width=\linewidth]{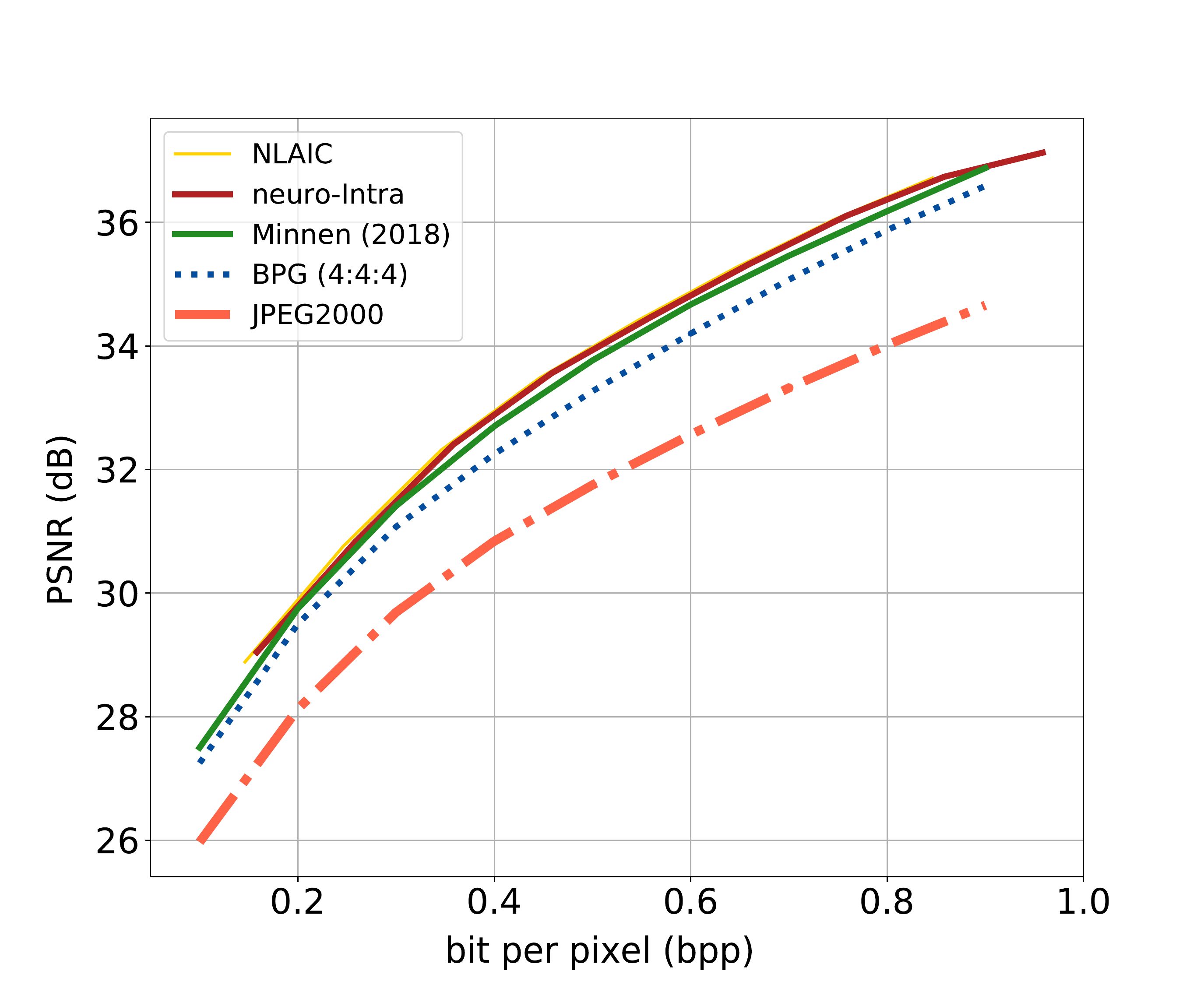}
	\caption{{\bf Efficiency of neuro-Intra.}  PSNR vs. rate performance of neuro-Intra in comparison to  NLAIC~\cite{chen2019neural}, Minnen (2018)~\cite{minnen2018joint}, BPG (4:4:4) and JPEG2000. Note that the curves for neuro-Intra and NLAIC overlap. }
	\label{fig:comp_of_universe_quantization}
\end{figure}

\subsection{Neural Intra Coding}
\label{sec:intra_coding}
Our neuro-Intra is a simplified version of the Non-Local Attention optimized Image Compression (NLAIC) that was originally proposed in \cite{chen2019neural}.

One major difference between the NLAIC and the VAE model using autoregressive spatial context in \cite{minnen2018joint} is the introduction of the NLAM inspired by~\cite{zhang2019residual}.  
In addition, we have applied 3D 5$\times$5$\times$5 masked CNN\footnote{This 5$\times$5$\times$5 convolutional kernel shares the same parameters for all channels, offering great model complexity reduction as compared with the 2D CNN-based solution in~\cite{minnen2018joint}.} to extract spatial priors, which are fused with hyper priors in PA for entropy context modeling ({\it e.g.}, the bottom part of Fig.~\ref{fig:entropy_model}).
Here, we have assumed the single Gaussian distribution for the context modeling of entropy coding. Note that temporal priors are not used for intra-pixel and inter-residual in this paper by only utilizing the spatial priors.

\begin{figure*}[t]
	\centering
{\includegraphics[scale=0.19]{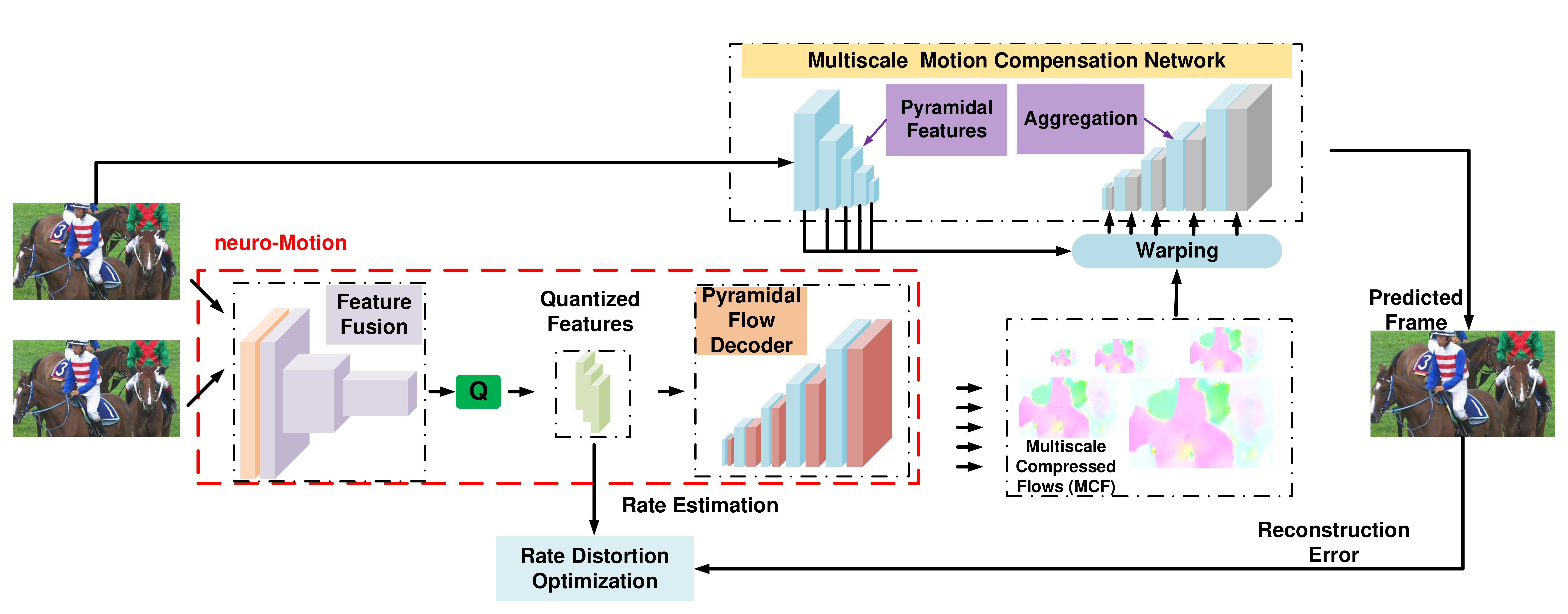}}
	\caption{{\bf Multiscale Motion Estimation and Compensation. }  One-stage neuro-Motion with MS-MCN uses a pyramidal flow decoder to synthesize the multiscale compressed optical flows (MCFs) that are used in a multiscale motion compensation network for generating predicted frames.}  \label{sfig:multi-scale}
\end{figure*}

The original NLAIC applies multiple NLAMs in both main and hyper coders, leading to excessive memory consumption at a large spatial scale. In E2E-NVC, NLAMs are only used at the bottleneck layers for both main and hyper encoder-decoder pairs, allowing bits to be allocated adaptively.

To overcome the non-differentiability of the quantization operation, quantization is usually simulated by adding uniform noise in~\cite{balle2018variational}. However, such noise augmentation is not exactly consistent with the rounding in inference, which can yield performance loss (as reported by~\cite{choi2019variable}). Thus, we apply  universal quantization (UQ)~\cite{choi2019variable} in neuro-Intra. UQ is used for neuro-Motion and neuro-Res as well. When applied to the common Kodak dataset, neuro-Intra performed as well as NLAIC~\cite{chen2019neural}, and outperformed Minnen (2018)~\cite{minnen2018joint}, BPG (4:4:4) and JPEG2000, as shown in Fig.~\ref{fig:comp_of_universe_quantization}.

\begin{figure}[t]
	\centering
	\includegraphics[scale=0.257]{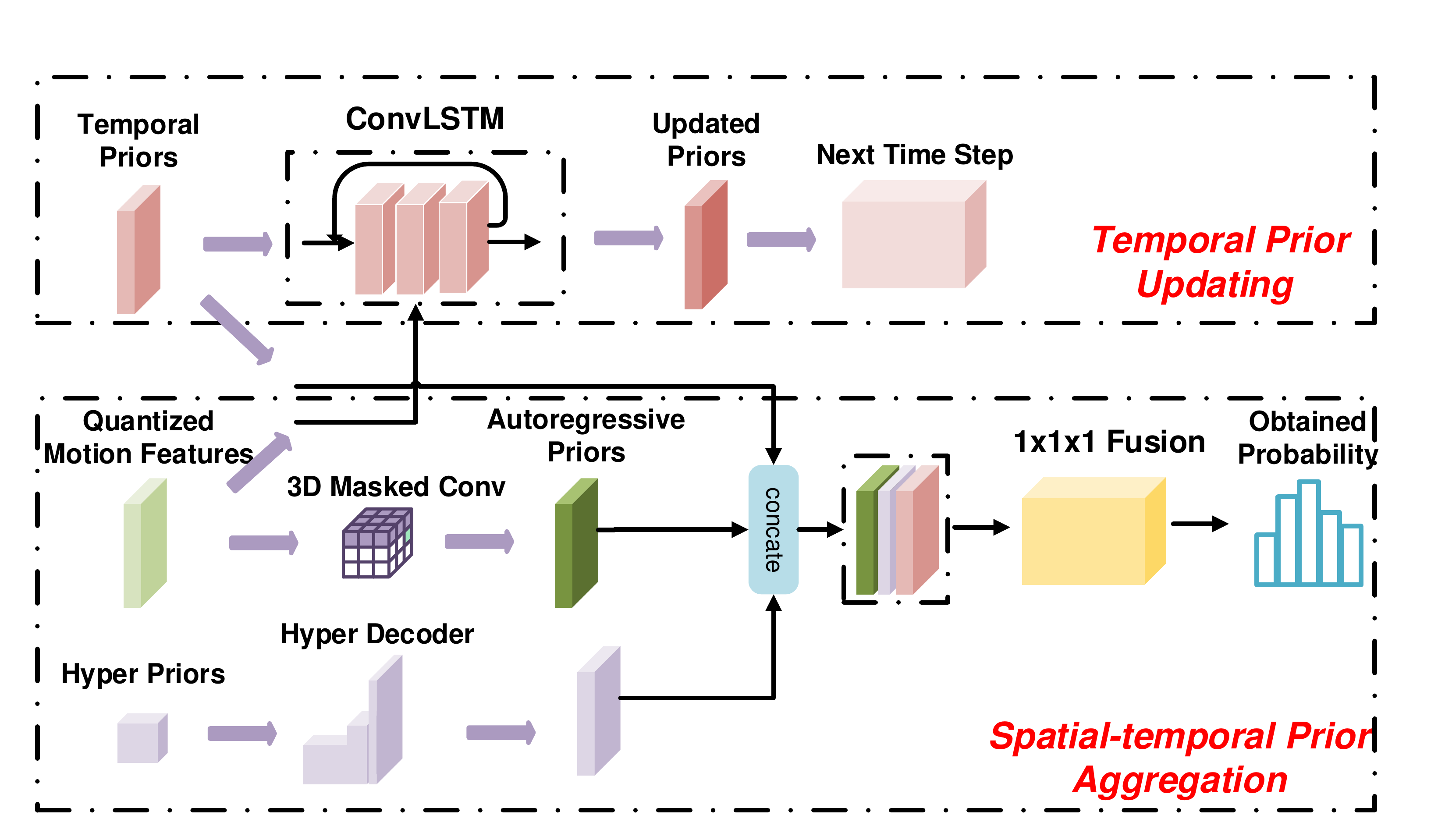}
	\caption{{\bf Context-Adaptive Modeling Using Joint Spatio-temporal and Hyper Priors.}  All priors are fused in PA to provide estimates of the probability distribution parameters. }
	\label{fig:entropy_model}
\end{figure}

\subsection{Neural Motion Coding and Compensation} \label{sec:inter_coding}
Inter-frame coding plays a vital role in video coding. The key is how to efficiently represent motion in a compact format for compensation. In contrast to the pixel-domain block-based motion estimation and compensation in conventional video coding, we rely on optical flow to accurately capture the temporal information for {\it motion compensation}.

To improve inter-frame prediction, we extend our earlier work~\cite{liu2020learned} to multiscale motion generation and compensation. 
This multiscale motion processing directly transforms  two concatenated frames (where one frame is the reference from the past, and one is the current frame) into quantized temporal features that represent the inter-frame motion. These quantized features are decoded into compressed optical flow in an unsupervised way for frame compensation via warping. This one-stage  scheme  does not require any pre-trained flow network such as FlowNet2 or PWC-net to generate the optical flow explicitly. It allows us to quantize the motion features rather than the optical flows, and to train the motion feature encoder and decoder together with explicit consideration of quantization and rate constraint.

The neuro-Motion module is modified for multiscale motion generation, where the main encoder is used for feature fusion. We replace the main decoder with  a {\it pyramidal flow decoder}, which generates the {multiscale compressed optical flows} (MCFs). MCFs  will be processed together with the reference frame, using a {\it multiscale  motion compensation} network (MS-MCN) to obtain the predicted frame efficiently, as shown in Fig.~\ref{sfig:multi-scale}.  Please refer to~\cite{liu2020neural} for more details.

Encoding motion compactly is another important factor for overall performance improvement. We suggest the joint spatio-temporal and hyper prior-based context-adaptive model shown in Fig.~\ref{fig:entropy_model} for efficiently inferring current quantized features. This is implemented in the PA engine of Fig.~\ref{fig:bio_video_coder}.

The joint spatio-temporal and hyper prior-based context-adaptive model mainly consists of  a {\it spatio-temporal-hyper aggregation module} (STHAM)  and a {\it temporal updating module} (TUM), shown in Fig.~\ref{fig:entropy_model}. At timestamp $t$, STHAM is introduced to accumulate all the accessible priors and estimate the mean and standard deviation of Gaussian Mixture Model (GMM) jointly using:
\begin{equation}
  (\mu_{\mathscr{F}},\sigma_{\mathscr{F}}) = \mathbb F( {\mathscr{F}}_1,..., {\mathscr{F}}_{i-1}, \hat{\bf z}_t, {\bf h}_{t-1}), \label{eq:flow_probability}
\end{equation}
Spatial priors are autoregressively derived using masked 5$\times$5$\times$5 3D convolutions and then concatenated with decoded hyper priors and temporal priors using stacked 1$\times$1$\times$1 convolutions. $\mathscr{F}_i, i = 0, 1, 2, ...$ are elements of quantized latent features ({\it e.g.}, motion flow), ${\bf h}_{t-1}$ is aggregated temporal priors from motion flows preceding the current frame. 
The neuro-Motion module exploits temporal redundancy to further prediction efficiency, leveraging the correlation between second-order moments of inter motion. A probabilistic model of each element to be encoded is derived with the estimated $\mu_{\mathscr{F}}$ and $\sigma_{\mathscr{F}}$ by:
\begin{align}
  &p_{{\mathscr{F}}|({\mathscr{F}}_1,..., {\mathscr{F}}{i-1},\hat{\bf z}_t, {\bf h}_{t-1})}({\mathscr{F}}_i|{\mathscr{F}}_1,..., {\mathscr{F}}_{i-1},\hat{\bf z}_t, {\bf h}_{t-1}) \nonumber\\
  & \mbox{~~~~~~~~~~~~~~~~}=  {\prod_i} (\mathcal{N}{(\mu_{\mathscr{F}},\sigma_{\mathscr{F}}^2)} *\mathcal{U}(-\frac{1}{2},\frac{1}{2})) ({\mathscr{F}}_i).
  \label{eq:residual_dist}
\end{align}

Note that TUM is applied to embedded current quantized features $\mathscr{F}_t$ recurrently using a standard ConvLSTM~\cite{shi2015convolutional}:
\begin{equation}
  ({\bf h}_t, {\bf c}_t) = {\rm ConvLSTM}({\mathscr{F}_t, {\bf h}_{t-1}, {\bf c}_{t-1}}),
  \label{eq:trn_priors}
\end{equation}
where ${\bf h}_t$ are updated temporal priors for the next frame, ${\bf c}_t$ is a memory state to control information flow across multiple time instances ({\it e.g.}, frames). Other recurrent units can also be used to capture  temporal correlations as in \eqref{eq:trn_priors}.

It is worth noting that leveraging second-order information for the representation of compact motion is also widely explored in traditional video coding approaches. For example, motion vector predictions from spatial and temporal co-located neighbors are standardized in H.265/HEVC, by which only motion vector differences (after prediction) are encoded.

\subsection{Neural Residual Coding} \label{ssec:residual_coding}
Inter-frame residual coding is another significant module contributing to the overall efficiency of the system. It is used to compress the temporal prediction error pixels.  It affects the efficiency of next frame prediction, since errors usually propagate temporally. 

Here we use the VAE architecture in Fig.~\ref{fig:bio_video_coder} to encode the  residual ${\bf r}_t$. The rate-constrained loss function is used:  
\begin{equation}
L = \lambda\cdot\mathbb{D}_2\left({\bf X}_t,({\bf X}^p_t+{\hat{\bf r}_t})\right) + R,
\label{eq:residual_generation}
\end{equation} where $\mathbb{D}_2$ is the $\ell_2$ loss between a residual compensated frame ${\bf X}^p_t+{\hat{\bf r}_t}$ and ${\bf X}_t$. neuro-Res will be first pretrained  using the frames predicted by the pretrained neuro-Motion and MS-MCN, and a loss function in~\eqref{eq:residual_generation} where the rate $R$  only accounts for the bits for residual. Then we refine neuro-Res jointly with neuro-Motion and MS-MCN, using a loss where $R$ incorporates the bits for both motion and residual with two frames. 

\subsection{Experimental Comparison}

\begin{figure}[h]
\renewcommand{\arraystretch}{1.06} 
\centering
\subfloat[]{\includegraphics[scale=0.4]{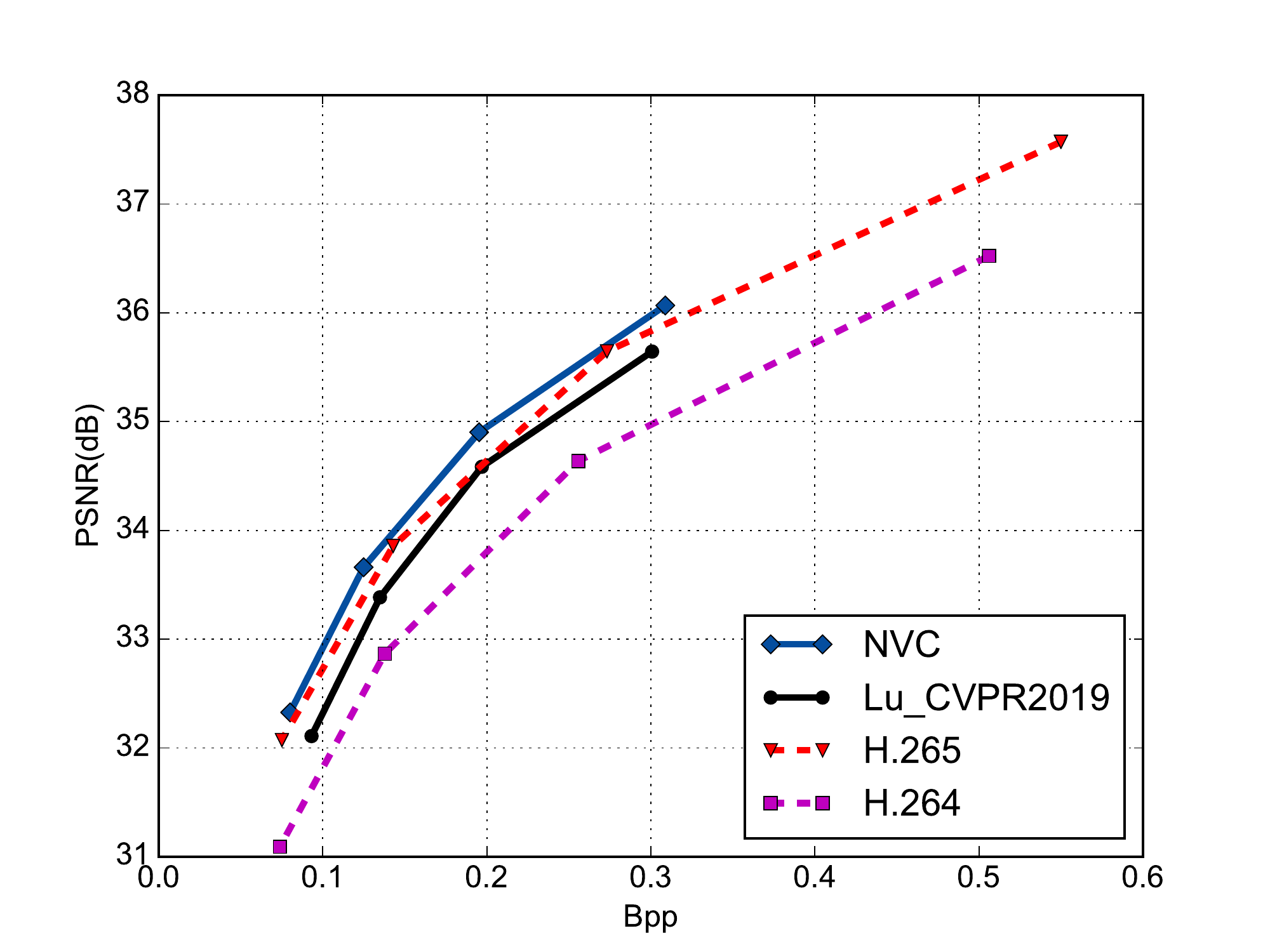}}\\
\subfloat[]{\includegraphics[scale=0.4]{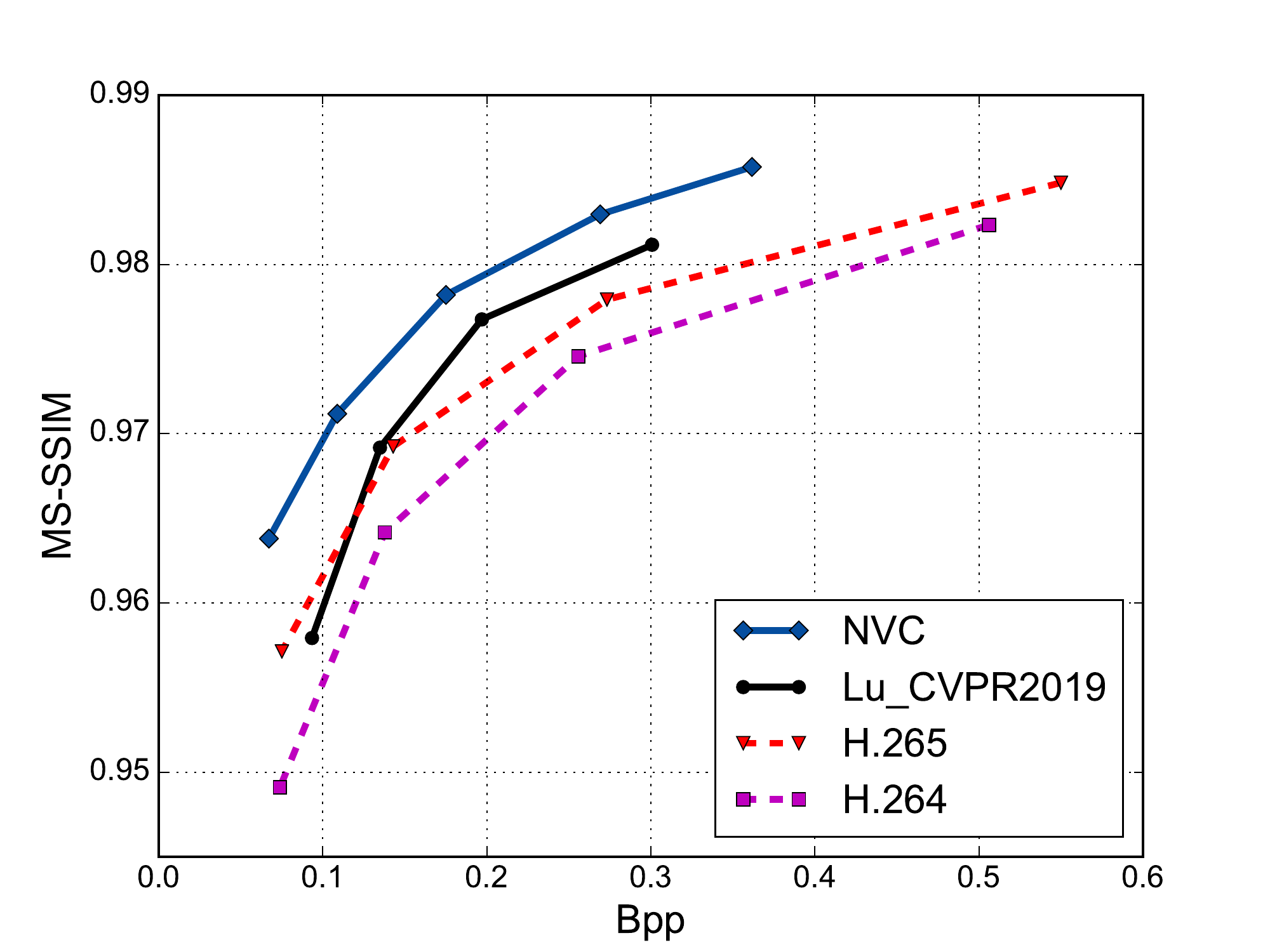}}
\caption{{\bf BD-Rate Illustration Using PSNR \& MS-SSIM.} (a)  NVC offers averaged 35.34\% gain against the anchor H.264/AVC when distortion is measured using PSNR. (b) NVC shows over 50\% gains against anchor H.264/AVC when using MS-SSIM evaluation. MS-SSIM is usually studied as a perceptual quality metric in image compression, especially at a low bit rate.
}
\label{rd_curve}
\end{figure}

\begin{table*}[t]
  \centering
  \caption{BD-Rate Gains of NVC, H.265/HEVC and DVC against the H.264/AVC. }
  \begin{tabular}{c|c|c|c|c|c|c|c|c|c|c|c|c}
    \hline
    \multirow{4}{*}{\textbf{Sequences}}&\multicolumn{4}{c|}{\textbf{H.265/HEVC}}&\multicolumn{4}{c|}{\textbf{DVC}}&\multicolumn{4}{c}{\textbf{NVC}}\\
     
    \cline{2-13}
      & \multicolumn{2}{c|}{PSNR} & \multicolumn{2}{c|}{MS-SSIM} &\multicolumn{2}{c|}{PSNR} & \multicolumn{2}{c|}{MS-SSIM} & \multicolumn{2}{c|}{PSNR} & \multicolumn{2}{c}{MS-SSIM} \\
      \cline{2-13}
    & {BDBR} & {BD-(D)} &{BDBR} & {BD-(D)} & {BDBR} & {BD-(D)}&{BDBR} &{BD-(D)} & {BDBR} & {BD-(D)} & {BDBR}& {BD-(D)}\\
    
     \hline
     
    {ClassB} & {-32.03\%} &{0.78} & {-27.67\%} & {0.0046} & {-27.92\%}& {0.72}& {-22.56\%}& {0.0049}& 
\bf{-45.66\%}& 
\bf{1.21}& 
\bf{-54.90\%}& 
\bf{0.0114}\\
    \hline
     {ClassC} & {-20.88\%} &{0.91} & {-19.57\%} & {0.0054} & {-3.53\%}& {0.13}& {-24.89\%}& {0.0081}& {-17.82\%}& 
\bf{0.73}& 
\bf{-43.11\%}& 
\bf{0.0133}\\
    \hline
    {ClassD} & {-12.39\%} &{0.57} & {-9.68\%} & {0.0023} & {-6.20\%}& {0.26}& {-22.44\%}& {0.0067}& 
\bf{-15.53\%}& 
\bf{0.70}& 
\bf{-43.64\%}& 
\bf{0.0123}\\
    \hline
    {ClassE} & {-36.45\%} &{0.99} & {-30.82\%} & {0.0018} & {-35.94\%}& {1.17}& {-29.08\%}& {0.0027}& 
\bf{-49.81\%}& 
\bf{1.70}& 
\bf{-58.63\%}& 
\bf{0.0048}\\
    \hline
    {UVG} & {-48.53\%} &{1.00} & {-37.5\%} & {0.0056} & {-37.74\%}& {1.00}& {-16.46\%}& {0.0032}& 
\bf{-48.91\%}& 
\bf{1.24}& 
\bf{-53.87\%}& 
\bf{0.0100}\\
    \hline
    {Average} & {-30.05\%} &{0.85} & {-25.04\%} & {0.0039} & {-22.26\%}& {0.65}& {-23.08\%}& {0.0051}& \bf{-35.54\%}& \bf{1.11}& \bf{-50.83\%}& \bf{0.0103}\\

    \hline
  \end{tabular}
  \label{tab:BDrate}
\end{table*}

\begin{figure*}[t]
     \centering
     \includegraphics[width=\linewidth]{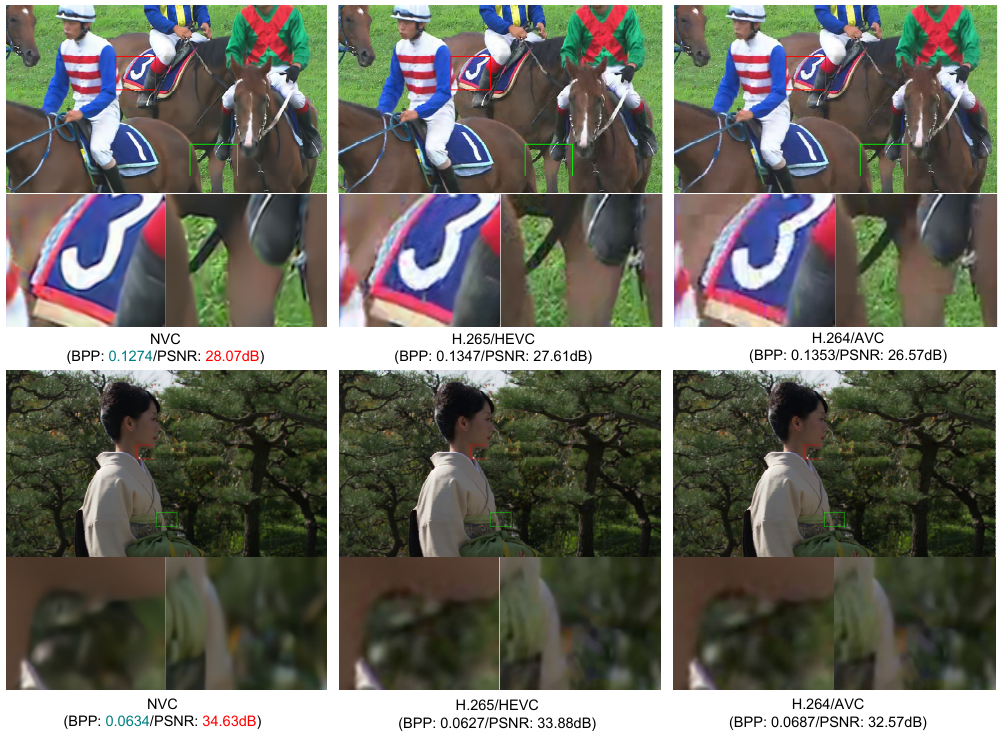}
     \caption{{\bf Visual Comparison.} Reconstructed frames of NVC, H.265/HEVC and H.264/AVC. We avoid blocky artifacts, visible noise, {\it etc.}, and provide better quality at lower bit rate.}
     \label{fig:visual_comparison}
\end{figure*}

We applied the same low-delay coding setting as DVC in~\cite{lu2019dvc} for our method and traditional
H.264/AVC, and H.265/HEVC for comparison. We encoded 100
frames and used GOP of 10 on H.265/HEVC test sequences, and 600
frames with GOP of 12 on the UVG dataset. For H.265/HEVC, we applied the fast mode of the x265\footnote{\url{http://x265.org/}} --- a popular open-source H.265/HEVC  encoder implementation; while the fast mode of the x264\footnote{\url{https://www.videolan.org/developers/x264.html}} is used as the representative of the H.264/AVC encoder.

We show the leading compression efficiency in  Fig.~\ref{rd_curve} using respective PSNR and MS-SSIM measures, across H.265/HEVC and UVG test sequences.
In Table~\ref{tab:BDrate}, by setting the same anchor using H.264/AVC, our NVC presents 35\% BD-Rate gains, while H.265/HEVC and DVC offer 30\% and 22\% gains, respectively. If the distortion is measured by the MS-SSIM, our gains in efficiency are even larger. This demonstrates that NVC can achieve a 50\% improvement in efficiency, while both H.265/HEVC and DVC achieve only around 25\%. 

Our NVC rivals the recent DVC\_Pro~\cite{lu2020end}, an upgrade of the earlier DVC~\cite{Lu_2019_CVPR}, {\it e.g.},   35.54\% and 50.83\% BD-Rate reduction measured by PSNR and MS-SSIM distortion respectively for NVC, while 34.57\% and 45.88\% marked for DVC\_Pro. DVC~\cite{Lu_2019_CVPR} has mainly achieved a higher level of coding efficiency than H.265/HEVC at high bit rates. However, a sharp decline in the performance of DVC is revealed at low bit rates ({\it e.g.}, performing worse than H.264/AVC at some rates). We have also observed that DVC's performance varies for different test sequences. DVC\_Pro upgrades DVC with better intra/residual coding using~\cite{minnen2018joint} and $\lambda$ fine-tuning, showing  state-of-the-art performance~\cite{lu2020end}.

{\bf Visual Comparison}
We provide a visual quality comparison between NVC, H.264/AVC, and H.265/HEVC as shown in Fig.~\ref{fig:visual_comparison}.  Generally,  NVC yields reconstructions that are much higher in quality than those of its competitors, even with a lower bit rate cost. For the sample clip ``RaceHorse'', which includes non-translational motion and a complex background, NVC uses 7 percent fewer bits despite an improvement in quality greater than 1.5 dB PSNR, compared with H.264/AVC. For other cases, our method also shows robust improvement.
Traditional codec usually suffers from blocky artifacts and motion-induced noise close to the edges of objects. In H.264/AVC, you clearly can observe block partition boundaries with severe pixel discontinuity. Our results provide higher-quality reconstruction and avoid noise and artifacts.

\subsection{Discussion And Future Direction}

We developed an end-to-end deep neural video coding framework that can learn compact spatio-temporal representation of raw video input. Our extensive simulations yielded very encouraging results, demonstrating that our proposed method can offer consistent and stable gains over existing methods ({\it e.g.}, traditional H.265/HEVC, recent learning-based approaches~\cite{lu2019dvc}, {\it etc.},) across a variety bit rates and a wide range of content.

The H.264/AVC, H.264/HEVC, AVS, AV1, and even the VVC, are masterpieces of hybrid prediction/transform framework-based video coding. Rate-distortion optimization, rate control, {\it etc.}, can certainly be incorporated to improve learning-based solutions. For example, reference frame selection is an important means by which we can embed and aggregate the most appropriate information for reducing temporal error and improving overall inter-coding efficiency. Making deep learning-based video coding practically applicable is another direction worthy of deeper investigation.

    \section{Case Studies for Post-processing:\\ Efficient Neural Filtering }
\label{sec:proposed_post_processing}

In this case study, both in-loop and post filtering are demonstrated using stacked DNN-based neural filters for quality enhancement of reconstructed frames. We specifically design a single-frame guided CNN which adapts pre-trained CNN models to different video contents for in-loop filtering, and a multi-frame CNN leveraging spatio-temporal information for post filtering. Both reveal noticeable performance gains.
In practice, neural filters can be devised, {\it i.e.}, in-loop or post, according to the application requirements.

\subsection{In-loop Filtering via Guided CNN}
\label{subsec:proposed_in_loop}

As reviewed in Section~\ref{sec:review_dnn_post_processing}, most existing works design a CNN model to directly map a degraded input frame to its restored version ({\it e.g.}, ground truth label), as illustrated in Fig.~\ref{fig:sec5_3_proposed_framework}a. To ensure that the model is generalizable to other contexts,  CNN models are often designed to use deeper layers, denser connections, wider receptive fields, {\it etc.}, with hundreds of millions of parameters. As a consequence, such generalized models are poorly suited to most practical applications. To address this problem, we propose that content adaptive weights be used to guide a shallow CNN model (as shown in Fig.~\ref{fig:sec5_3_proposed_framework}b) instead. 

The principle underlying this approach is sparse signal decomposition: We expect that the CNN model can represent any input as a weighted combination of channel-wise features. Note that weighting coefficients are dependent on input signals, making this model generalizable to a variety of content characteristics.

{\bf{Method.}}
Let $\bf{x}$ be a degraded block with $\rm N$ pixels in a column-wise vector format. The corresponding source block of $\bf{x}$ is $\bf {s}$, which has a processing error  $\bf d=s-x$. We wish to have ${\bf r}_{\rm corr}$ from $\bf {x}$ so that the final reconstruction ${\bf x}_{\rm corr} = {\bf x} + {\bf r}_{\rm corr}$ is closer to $\bf s$. 

\begin{figure}[t]
	\centering
	\centerline{\includegraphics[width=0.9\linewidth]{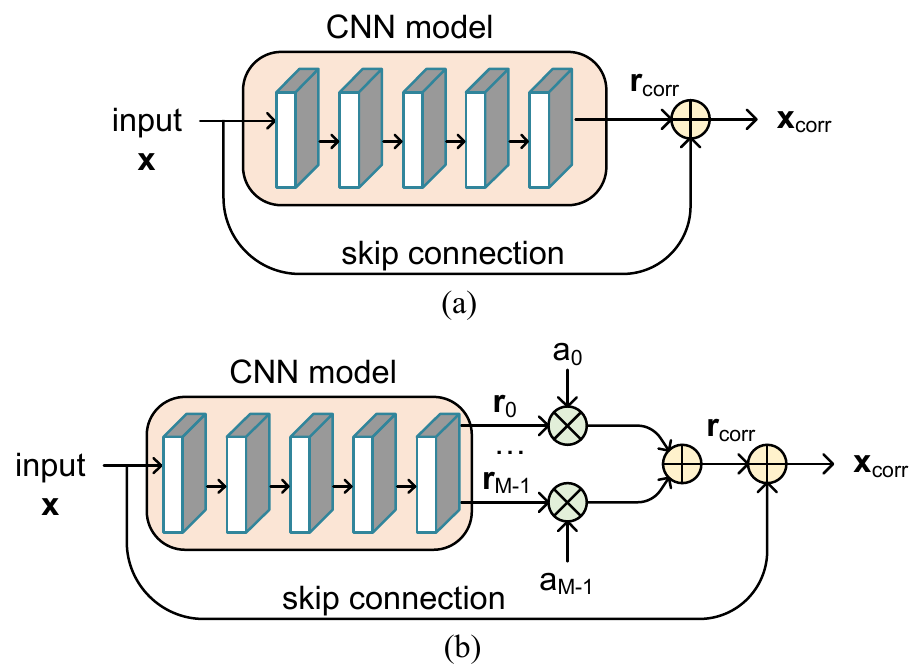}}
	\caption{{\bf CNN-based Restoration.} (a) Conventional model structure. (b) Guided CNN model with adaptive weights.}
	\label{fig:sec5_3_proposed_framework}
\end{figure}

Let the CNN output layer have $\rm M$ channels, {\it i.e.}, ${\bf r}_{\rm 0}$, ${\bf r}_{\rm 1}$,$\cdots$,${\bf r}_{\rm M-1}$. Then, the ${\bf r}_{\rm corr}$ is assumed as a linear combination of these channel-wise feature vectors,
\begin{equation}
\label{eq:cal_a}
{\bf r}_{\rm corr} = {a_0}{\bf r}_{\rm 0} + {a_1}{\bf r}_{\rm 1}+\cdots+{a_{\rm M-1}}{\bf r}_{\rm M-1},
\end{equation}
where $a_0,a_1,\cdots,a_{\rm M-1}$ are the weighting parameters that are explicitly signaled in the compressed bitstream.  

Our objective is to minimize the distance between the restored block ${\bf x}_{\rm corr}$ and its corresponding source $\bf s$, {\it i.e.}, $\left|{\bf x}_{\rm corr} -{\bf s}\right|^2 = \left|{\bf r}_{\rm corr} -{\bf d}\right|^2$. Given the channel-wise output features $\bf r_{\rm 0}$, $\bf r_{\rm 1}$, $\cdots$, $\bf r_{\rm M-1}$, for a degraded input $\bf x$, the weighting parameters $a_0,a_1,\cdots,a_{\rm M-1}$ can then be estimated by least-square optimization as
\begin{equation}
\left[a_0,a_1,\cdots,a_{\rm M-1}\right]^{\rm T} = ({\bf R}^{\rm T}{\bf R})^{-1}{\bf R}^{\rm T}{\bf d},
\end{equation}
where ${\bf R}=\left[{\bf r}_0, {\bf r}_1,\dots,{\bf r}_{\rm M-1}\right]$ is the matrix at a size of $\rm N\times M$ comprised of stacked output features in column-wise order.
The reconstruction error is given by
\begin{equation}
e=|{\bf r_{\rm corr}} -{\bf d}|^2 = |{\bf d}|^2-{\bf d}^{\rm T} {\bf R}({\bf R}^{\rm T}{\bf R})^{-1}{\bf  R}^{\rm T}{\bf d}.
\end{equation}

{\bf Loss Function.}
Assuming that one training batch is comprised of $\rm T$ patch pairs: $\{{\bf s}_i,{\bf x}_i\}, i=0,1,,\cdots,\rm T-1$,  the overall reconstruction error over the training set is
\begin{equation}
E=\sum\nolimits_i {\{ {{\left| {{{\bf{d}}_i}} \right|}^2} - {{\bf{d}}_i}^{\rm{T}}{{\bf{R}}_i}{{({{\bf{R}}_i}^{\rm{T}}{{\bf{R}}_i})}^{ - 1}}{{\bf{R}}_i}^{\rm{T}}{{\bf{d}}_i}\} },
\end{equation}
where ${\bf d}_i={\bf s}_i-{\bf x}_i$ is the error for the $i^{th}$ patch. ${\bf R}_i=[{\bf r}_{i,0},{\bf r}_{i,1},\cdots,{\bf r}_{i,{\rm M-1}}]$ is the corresponding channel-wise features in matrix form, with ${\bf r}_{i,j}$ being the $j^{th}$ channel when training sample $\bf x_i$ is passed through the CNN model. Given that $\left|{\bf d}_i\right|^2$ is independent of the network model, the loss function can be simplified as 
\begin{equation}
L=\sum\nolimits_i {\{ - {{\bf{d}}_i}^{\rm{T}}{{\bf{R}}_i}{{({{\bf{R}}_i}^{\rm{T}}{{\bf{R}}_i})}^{ - 1}}{{\bf{R}}_i}^{\rm{T}}{{\bf{d}}_i}\} }.
\end{equation}

{\bf Experimental Studies.}  A shallow baseline CNN model(as described in Table~\ref{tab:proposed_cnn_structure}) is used to demonstrate the efficiency of the guided CNN model. This model is comprised of seven layers in total and has a fixed kernel size of 3$\times$3. At the bottleneck layer, the channel number of the output feature map is $\rm M$. After extensive simulations, $\rm M=2$ was selected. In total, our model only requires 3,744 parameters, far fewer than the number required by existing methods.

\begin{table}[]
	\renewcommand{\arraystretch}{1.06}
	\caption{Layered structure and parameter settings of baseline CNN model.}
	\label{tab:proposed_cnn_structure}
	\centering
	\resizebox{\linewidth}{!}
	{
		\begin{tabular}{ccSSS} \hline 
			Layer & Kernel size & {Input channels}  & {Output channels}   & {Parameters} \\ \hline
			1     & $3\times3$  & 1  & 16 &  144   \\ 
			2     & $3\times3$  & 16 & 8  &  1152  \\ 
			3     & $3\times3$  & 8  & 8  &  576  \\ 
			4     & $3\times3$  & 8  & 8  &  576  \\
			5     & $3\times3$  & 8  & 8  &  576  \\
			6     & $3\times3$  & 8  & 8  &  576  \\
			7     & $3\times3$  & 8 & $\rm M=2$ & 144 \\ \hline
			\multicolumn{4}{c}{Total parameters} & 3744 \\ \hline
		\end{tabular}
	}
\end{table}

In training, 1000 pictures of DIV2K~\cite{DIV2K} were used. All frames were compressed using the AV1 encoder with in-loop filters CDEF ~\cite{midtskogen2018av1} and LR~\cite{mukherjee2017switchable} turned off to generate corresponding quantization-induced degraded reconstructions. We divided the 64 QPs into six ranges and trained one model for each QP range. The six ranges include QP values 7 to 16, 17 to 26, 27 to 36, 47 to 56, and 57 to 63.  Compressed frames falling into the same QP range were used to train the corresponding CNN model. Frames were segmented into 64$\times$64 patches. Each batch contained 1,000 patches. We adopted the Adaptive moment estimation (Adam) algorithm, with the initial learning rate set at 1e-4. The learning rate is halved every 20 epochs. 

We used the Tensorflow platform, which runs on NVIDIA GeForce GTX 1080Ti GPU, to evaluate coding efficiency across four QPs, {\it e.g.}, \{32, 43, 53, and 63\}. Our test set included 24 video sequences with resolutions ranging from 2560$\times$1600 to 352$\times$288. The first 50 frames of each sequence were tested in both intra and inter configurations. 

In our experiments, $\rm N$ was set to 64, 128, 256, and the whole frame, respectively. We found that $\rm N=256$ yields the best performance. For each block, the linear combination parameters $a_i$ $(i=0,1)$ were derived accordingly. To strike an appropriate balance between bit consumption and model efficiency, our experiments suggest that the dynamic range of $a_i$ is within 15.

We compared the respective BD-Rate reductions of our guided CNN model and a baseline CNN model against the AV1 baseline encoder. All filters were enabled for the AV1 anchor. For a description of the baseline CNN model, see Table~\ref{tab:proposed_cnn_structure}. Our guided CNN model is the baseline model plus the adaptive weights. 

Both baseline and guided CNN models were applied on top of the AV1 encoder with only the deblocking filter enabled, and other filters (including CDEF and LR) turned off. The findings reported in Table~\ref{tab:BD-rate_performance} demonstrate that either baseline or guided CNN models can be used to replace additional adaptive in-loop filters, while improving R-D efficiency. Furthermore, regardless of block size and frame types, our guided model always outperformed the baseline CNN. This is mainly due to the adaptive weights used to better characterize content dynamics.  Similar lightweight CNN structures can be upgraded using deep models~\cite{RHCNN, VRCNN,MMS-net} for potentially greater BD-Rate savings.

\begin{table*}[] 
	\renewcommand{\arraystretch}{1.06}
	\caption{BD-Rate savings of baseline and guided CNN models against the AV1.}
	\label{tab:BD-rate_performance}	
	\centering
	\resizebox{\linewidth}{!}
	{
		\begin{tabular}{cc|d{2.3}|d{2.3}d{2.3}d{2.3}d{2.3}|d{2.3}|d{2.3}d{2.3}d{2.3}d{2.3}} \hline 
			\multirow{3}{*}{Resolution}  & \multirow{3}{*}{Sequence}     & \multicolumn{5}{c|}{All Intra}  & \multicolumn{5}{c}{Random Access} \\ \cline{3-12}
			&   & \multicolumn{1}{c|}{Baseline} & \multicolumn{4}{c|}{Guided CNN}   & \multicolumn{1}{c|}{Baseline} & \multicolumn{4}{c}{Guided CNN}      \\
			&   & \multicolumn{1}{c|}{CNN}  &  \multicolumn{1}{c}{N=64}     &  \multicolumn{1}{c}{N=128}    &  \multicolumn{1}{c}{N=256}    & \multicolumn{1}{c|}{Frame}   & \multicolumn{1}{c|}{CNN}  &  \multicolumn{1}{c}{N=64}     &  \multicolumn{1}{c}{N=128}    &  \multicolumn{1}{c}{N=256}    & \multicolumn{1}{c}{Frame}   \\ \hline
			\multirow{2}{*}{ $2560\times1600$ }               
			& PeopleOnStreet  &  -1.15\%    &  -1.95\%  &  -2.84\%  &  -2.90\%  &  -2.81\%  &  -0.19\%   &  -0.22\%  &  -1.03\%  &  -1.02\%  &  -0.83\%  \\
			& Traffic         &  -1.71\%    &  -1.76\%  &  -3.01\%  &  -3.16\%  &  -3.03\%  &  -0.26\%   &  +1.89\%   &  -1.64\%  &  -2.15\%  &  -2.17\%  \\
			\hline
			\multirow{7}{*}{ $1920\times1080$ } 
			& BasketballDrive &  -0.45\%  &  +2.95\%  &  -0.72\%  &  -1.06\%  &  -0.72\%  &  -0.02\%  &  +8.04\%  &  +0.87\%  &  +0.07\%  &  -0.05\%  \\
			& BQTerrace       &  -0.98\%  &  -3.19\%  &  -3.66\%  &  -3.44\%  &  -2.10\%  &  -0.33\%  &  +0.68\%  &  -1.62\%  &  -1.91\%  &  -1.51\%  \\
			& Cactus          &  -1.64\%  &  -1.38\%  &  -2.79\%  &  -2.89\%  &  -2.56\%  &  -0.21\%  &  +1.18\%  &  -1.13\%  &  -1.31\%  &  -0.96\%  \\
			& Kimono          &  -0.23\%  &  +3.55\%  &  -0.18\%  &  -0.88\%  &  -0.95\%  &  -0.07\%  &  +6.07\%  &  +0.84\%  &  -0.07\%  &  -0.01\%  \\
			& ParkScene       &  -1.21\%  &  +0.01\%  &  -1.92\%  &  -2.21\%  &  -2.11\%  &  -0.07\%  &  +1.11\%  &  -1.46\%  &  -1.82\%  &  -0.92\%  \\
			& blue-sky        &  -2.89\%  &  -0.96\%  &  -2.58\%  &  -2.86\%  &  -2.56\%  &  +0.00\%  &  +3.46\%  &  -2.02\%  &  -2.96\%  &  -2.77\%  \\
			& crowd\_run      &  -3.01\%  &  -2.34\%  &  -3.11\%  &  -3.22\%  &  -3.08\%  &  -0.13\%  &  -1.69\%  &  -2.19\%  &  -2.07\%  &  -1.09\%  \\
			\hline
			\multirow{4}{*}{ $832\times480$ }   
			& BasketballDrill &  -2.99\%    &  -5.55\%  &  -6.45\%  &  -6.26\%  &  -5.88\%  &  -0.25\%  &  -0.33\%  &  -2.10\%  &  -1.79\%  &  -1.55\%  \\
			& BQMall          &  -1.74\%    &  -3.96\%  &  -4.48\%  &  -4.46\%  &  -4.35\%  &  -0.15\%  &  +0.16\%  &  -1.05\%  &  -1.13\%  &  -0.76\%  \\
			& PartyScene      &  -0.83\%    &  -3.77\%  &  -4.02\%  &  -3.97\%  &  -3.81\%  &  -0.20\%  &  -1.10\%  &  -1.43\%  &  -1.25\%  &  -0.13\%  \\
			& RaceHorsesC     &  -1.91\%    &  -2.01\%  &  -2.58\%  &  -2.49\%  &  -2.38\%  &  -0.21\%  &  -0.70\%  &  -1.28\%  &  -1.03\%  &  -0.80\%  \\
			\hline
			\multirow{4}{*}{ $416\times240$ }   
			& BasketballPass  &  -3.08\%    &  -3.66\%  &  -4.60\%  &  -4.72\%  &  -4.65\%  &  -0.20\%  &  +0.71\%  &  -0.63\%  &  -0.62\%  &  -0.36\%  \\
			& BlowingBubbles  &  -2.60\%    &  -3.36\%  &  -3.78\%  &  -3.77\%  &  -3.76\%  &  -0.34\%  &  -0.55\%  &  -1.05\%  &  -0.87\%  &  -0.86\%  \\
			& BQSquare        &  -4.92\%    &  -6.09\%  &  -6.23\%  &  -6.27\%  &  -6.22\%  &  -0.50\%  &  -0.54\%  &  -0.92\%  &  -1.13\%  &  -1.17\%  \\
			& RaceHorses      &  -3.57\%    &  -5.39\%  &  -5.75\%  &  -5.75\%  &  -5.76\%  &  -0.51\%  &  -2.82\%  &  -3.06\%  &  -2.69\%  &  -2.94\%  \\
			\hline
			\multirow{3}{*}{ $1280\times720$ }  
			& Johnny          &  -2.01\%    &  -2.41\%  &  -4.03\%  &  -4.21\%  &  -4.12\%  &  -0.31\%  &  +8.32\%   &  -0.94\%  &  -2.57\%  &  -2.63\%  \\
			& FourPeople      &  -1.94\%    &  -0.54\%  &  -3.49\%  &  -3.76\%  &  -2.85\%  &  -0.29\%  &  +17.99\%  &  +1.20\%  &  -1.65\%  &  -1.60\%  \\
			& KristenAndSara  &  -2.71\%    &  -1.49\%  &  -3.97\%  &  -4.32\%  &  -4.26\%  &  -0.42\%  &  +15.95\%  &  +0.53\%  &  -2.49\%  &  -2.31\%  \\
			\hline
			\multirow{4}{*}{ $352\times288$ }   
			& Harbour         &  -0.79\%    &  -1.18\%  &  -1.43\%  &  -1.38\%  &  -1.42\%  &  -0.23\%  &  -1.00\%  &  -1.29\%  &  -1.40\%  &  -1.08\%  \\
			& Ice             &  -3.59\%    &  -5.54\%  &  -6.88\%  &  -7.08\%  &  -7.19\%  &  -0.59\%  &  -1.59\%  &  -3.59\%  &  -3.65\%  &  -3.97\%  \\
			& Silent          &  -1.68\%    &  -1.88\%  &  -2.80\%  &  -2.77\%  &  -2.79\%  &  -0.21\%  &  +1.96\%  &  -0.29\%  &  -0.27\%  &  -0.70\%  \\
			& Students        &  -3.08\%    &  -4.10\%  &  -4.77\%  &  -4.81\%  &  -4.88\%  &  -0.52\%  &  +1.25\%  &  -1.16\%  &  -1.44\%  &  -1.66\%  \\ \hline
			& {Average} &  -2.11\%  &  -2.33\%  &  -3.59\%  &  -3.69\%  &  -3.51\%  &  -0.26\%  &  +2.43\%  &  -1.10\%  &  -1.55\%  &  -1.37\%  \\ \hline
		\end{tabular}
	}
\end{table*}

\subsection{Multi-frame Post Filtering}
\label{subsec:proposed_out_loop}

This section demonstrates how multi-frame video enhancement (MVE) scheme-based post filtering can be used to minimize compression artifacts. We implemented our proposed approach on AV1 reconstructed frames and achieved significant coding improvement. 
Similar observations are expected with different anchors, such as the H.265/HEVC.

{\textbf{Method.}} Single-frame video enhancement (SVE) refers to the sole application of the fusion network without leveraging temporal frame correlations. 
As discussed in Section~\ref{sec:review_dnn_post_processing}, there are a great number of network models that can be used to do SVE. In most cases, the efficiency and complexity are at odds with one another: In other words, efficiency and complexity come at the cost of deeper networks and higher numbers of parameters. Recently, Yu \etal~\cite{WDSR} discovered that models with more feature channels before activation could provide significantly better performance with the same parameters and computational budgets. We designed a wide activation residual network (WARN) by combining wide activation with a powerful deep residual network (ResNet)~\cite{ResNet}, shown in Fig.~\ref{fig:sec5_3_WARN_structure}. This WARN illustrates the three inputs for an enhanced output in the MVE framework. In contrast, SVE normally inputs a single frame, and outputs a corresponding enhanced representation.

\begin{figure}[!t]
	\centering
	\centerline{\includegraphics[width=\linewidth]{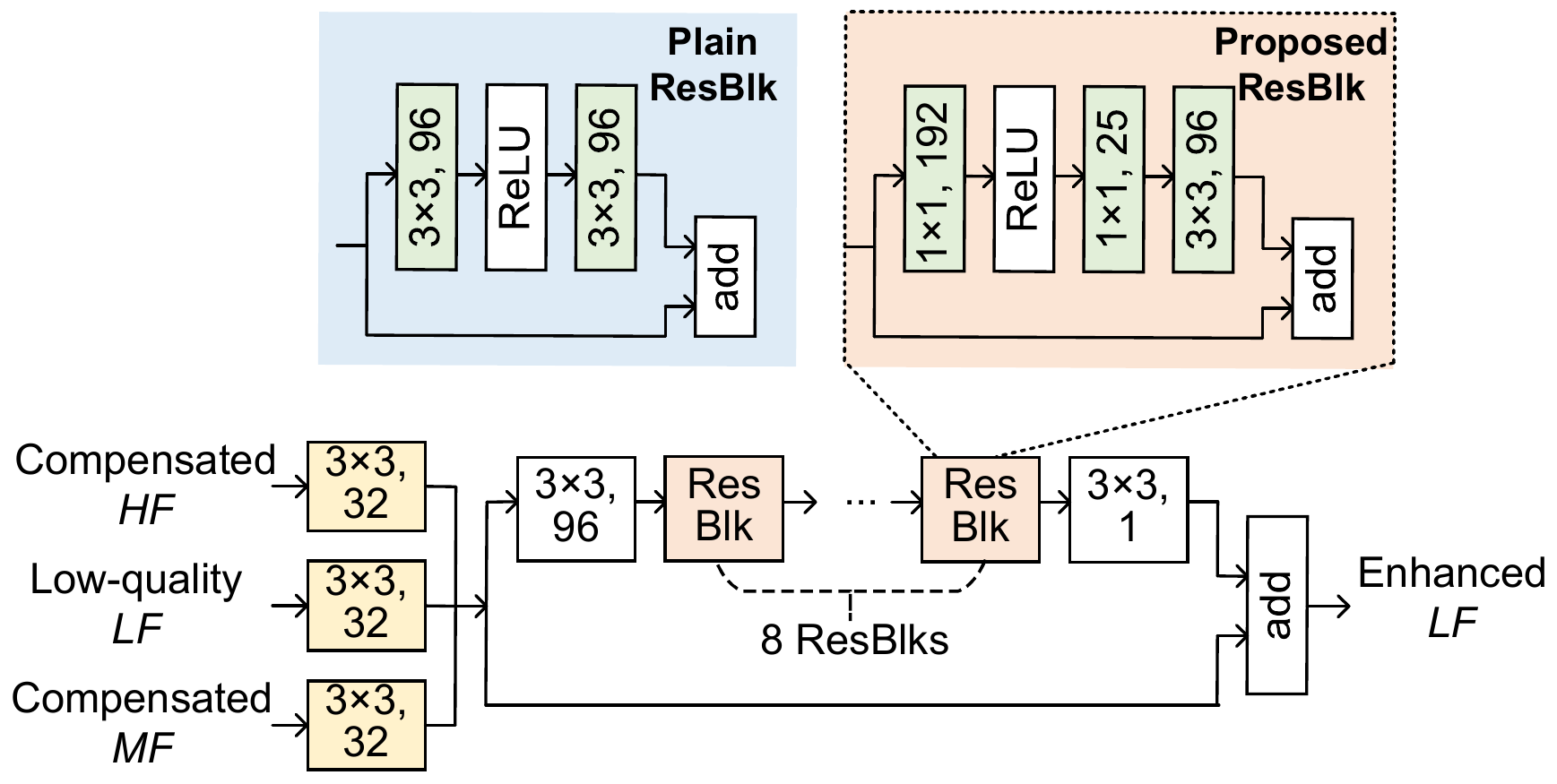}}
	\caption{{\bf WARN.} This wide activation residual network is used to fuse/enhance input frame for improved quality. In MVE case, it takes three inputs to enhance the LFs; and in SVE case, it inputs a single frame and outputs its enhanced version. This WARN generally follows the residual network structure with residual link and ResBlk embedded. Note that  ResBlk is extended to support wide activation from its plain version prior to ReLU activation. }
	\label{fig:sec5_3_WARN_structure}
\end{figure}

This MVE closely follows the two-step strategy reviewed in Section~\ref{sec:review_dnn_post_processing}. It uses FlowNet2~\cite{ilg2017flownet} to perform pixel-level motion estimation/compensation-based temporal frame alignment. Next, a WARN-based fusion network is used for final enhancement. We allow the two High-quality Frames (HF) immediately preceding and succeeding a low-quality frame (LF) to enhance the Low-quality Frame (LF) in between. Bi-directional warping is performed for each LF to produce compensated HFs  in Fig.~\ref{fig:sec5_3_MVE_framework}.

\begin{figure*}[!t]
	\centering
	\centerline{\includegraphics[width=\linewidth]{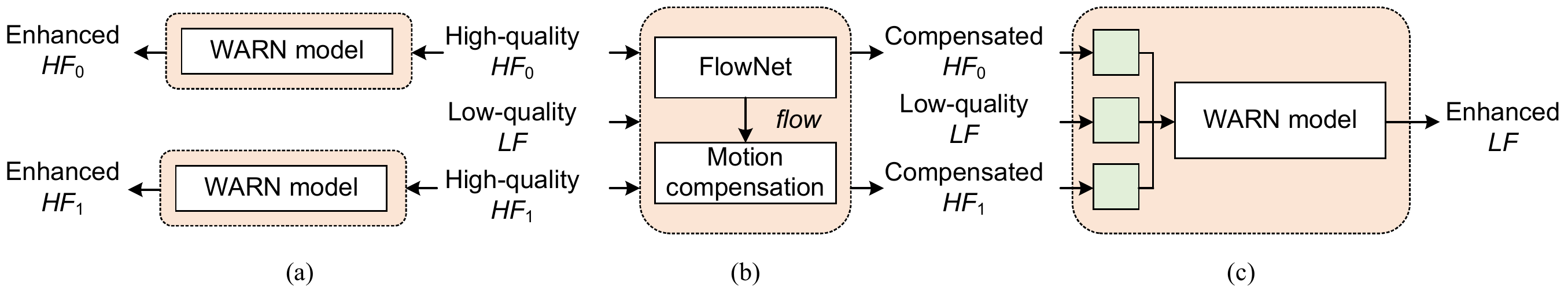}}
	\caption{{\bf Enhancement Framework.} (a) Single-input WARN-based SVE to enhance the HF. (b)+(c) Two-step MVE using FlowNet2 for temporal alignment, and three-input WARN- based fusion to use preceding and succeeding HFs for LF enhancement.}
	\label{fig:sec5_3_MVE_framework}
\end{figure*}

{\textbf{Experimental Studies.}}
We evaluate both SVE and MVE against the AV1 baseline. A total of 118 video sequences were selected to train network models. More specifically, the first 200 frames of each sequence were encoded with AV1 encoder to generate the reconstructed frames. The QPs are \{32, 43, 53, 63\}, yielding 23,600 reconstructed frames in total. 
After frame alignment, we selected one training set containing compensated $HF_0$, compensated $HF_1$, and to-be-enhanced LF  from every 8 frames, which yielded a total of 2900 training sets. These sets were used to train the WARN model as the fusion network. 
Notice that we trained the WARN models for SVE and MVE individually. 
The GoP size was 16 with a hierarchical prediction structure. The LFs and HFs were identified using their QPs, {\it i.e.}, HFs with lower QP than the base QP were decoded, such as  frames 0, 4, 8, 12, and 16 in Fig.~\ref{fig:sec7_loop_filter_hierarchical_AV1}. 

Algorithms were implemented using the Tensorflow platform,  NVIDIA GeForce GTX 1080Ti GPU. In training, frames were segmented into 64$\times$64 patches, with 64 patches included in each batch.  We adopted the Adam optimizer with the initial learning rate set at 1e{-4}. The learning rate can be then adjusted using the step strategy with $\gamma = 0.5$. 
An additional 18 sequences were also employed for testing. These were mostly used to evaluate video quality. The first 50 frames of each test sequence were compressed. Then the reconstructed frames were enhanced using the proposed SVE and MVE methods. 

We applied the proposed method on AV1 reconstructed frames. The results are presented in Table~\ref{tab:compared_with_AV1_and_single}. Due to the hierarchical coding structure in inter prediction, the LFs in Fig. \ref{fig:sec7_loop_filter_hierarchical_AV1} were enhanced using the neighboring HFs via MVE framework. The HFs themselves are enhanced using the SVE method. 

\begin{figure}[t]
	\centering
	\centerline{\includegraphics[width=\linewidth]{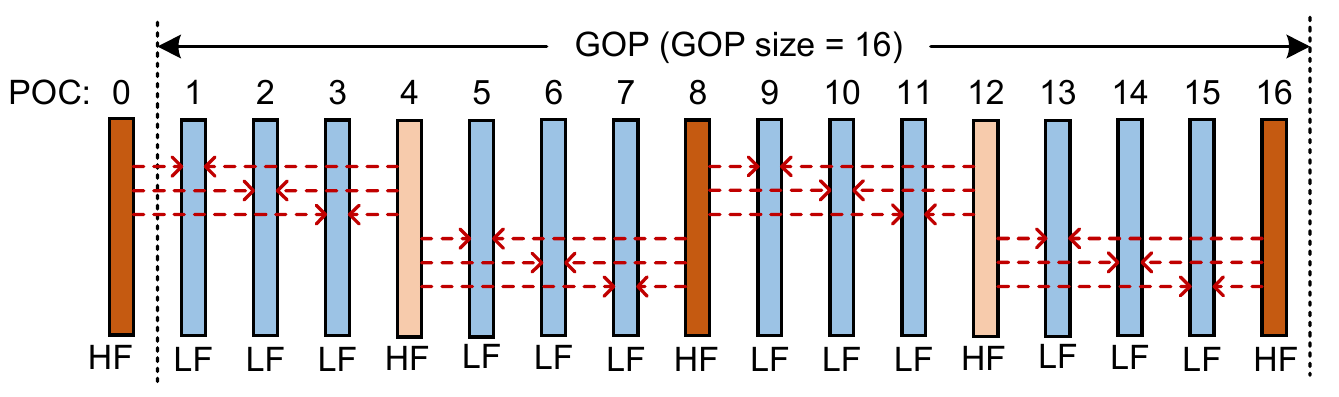}}
	\caption{{\bf The hierarchical coding structure in the AV1 encoder}. The  LFs are enhanced using HFs following the prediction structure via MVE scheme, and HFs are restored using SVE method.}
	\label{fig:sec7_loop_filter_hierarchical_AV1}
\end{figure}

The overall BD-Rate savings of the SVE and MVE methods are tabulated in Table~\ref{tab:compared_with_AV1_and_single}, against the AV1. SVE achieves an averaged reduction of 8.2\% and 5.0\% BD-rate for all intra and random access scenarios, respectively. On the other hand, our MVE obtains 20.1\% and 7.5\% BD-rate savings on average, further demonstrating the effectiveness of our proposed scheme. When random access techniques are used, the HFs selected are generally distant from a target LF, which reduces the benefits provided from inter HFs. On the other hand, intra coding techniques uniformly demonstrate greater BD-rate savings, because the neighboring frames nearest to target LFs can be used. This contributes significantly to enhancement.

Besides the objective measures, sample snapshots of reconstructed frames are illustrated in Fig.~\ref{fig:sec7_loop_filter_perceptual_comparison}, clearly demonstrating that blocky and ringing  artifacts from the AV1 baseline are attenuated after applying either SVE or MVE based filtering. Notably, MVE creates more visually appealing images than SVE. 

\begin{table}[]
	\renewcommand{\arraystretch}{1.06}
	\caption{BD-rate improvement of proposed SVE and MVE scheme against the AV1.}
	\label{tab:compared_with_AV1_and_single}	
	\centering
	\resizebox{\linewidth}{!}
	{
		\begin{tabular}{cc|d{2.2}d{2.2}|d{2.2}d{2.2}} \hline
			\multirow{2}{*}{Class} & \multirow{2}{*}{Sequence} & \multicolumn{2}{c|}{All Intra}  & \multicolumn{2}{c}{Random Access}  \\
			&                           & \multicolumn{1}{c}{SVE} & \multicolumn{1}{c|}{MVE} & \multicolumn{1}{c}{SVE} & \multicolumn{1}{c}{MVE} \\ \hline
			\multirow{2}{*}{A}     
			& PeopleOnStreet            & -9.1\%      & -14.7\%    & -5.0\%      & -8.1\%     \\
			& Traffic                   & -7.6\%      & -22.2\%    & -5.8\%      & -8.8\%     \\ \hline
			\multirow{5}{*}{B}     
			& BasketballDrive           & -5.9\%      & -13.1\%    & -4.4\%      & -6.4\%     \\
			& BQTerrace                 & -8.0\%      & -23.7\%    & -7.7\%      & -9.8\%     \\
			& Cactus                    & -7.7\%      & -21.9\%    & -3.9\%      & -6.0\%     \\
			& Kimono                    & -3.8\%      & -20.4\%    & -3.9\%      & -7.1\%     \\
			& ParkScene                 & -5.1\%      & -26.3\%    & -4.9\%      & -8.0\%     \\ \hline
			\multirow{4}{*}{C}    
			& BasketballDrill           & -12.5\%     & -21.3\%    & -5.6\%      & -7.9\%     \\
			& BQMall                    & -8.9\%      & -18.7\%    & -3.5\%      & -6.1\%     \\
			& PartyScene                & -7.2\%      & -19.0\%    & -3.2\%      & -5.0\%     \\
			& RaceHorsesC               & -5.9\%      & -18.3\%    & -3.3\%      & -5.6\%     \\ \hline
			\multirow{4}{*}{D}     
			& BasketballPass            & -10.0\%     & -18.5\%    & -3.4\%      & -6.2\%     \\
			& BlowingBubbles            & -7.0\%      & -19.8\%    & -4.6\%      & -6.7\%     \\
			& BQSquare                  & -10.8\%     & -21.3\%    & -11.0\%     & -13.6\%    \\
			& RaceHorses                & -9.2\%      & -19.3\%    & -4.9\%      & -7.8\%     \\ \hline
			\multirow{3}{*}{E}     
			& FourPeople                & -9.7\%      & -21.7\%    & -5.1\%      & -7.4\%     \\
			& Johnny                    & -9.6\%      & -20.7\%    & -5.5\%      & -8.0\%     \\
			& KristenAndSara            & -9.6\%      & -21.2\%    & -4.4\%      & -7.0\%     \\ \hline
			& Average                   & -8.2\%      & -20.1\%    & -5.0\%      & -7.5\%     \\
			\hline 
		\end{tabular}
	}
\end{table}

\begin{figure}[!t]
	\centering
	\centerline{\includegraphics[width=\linewidth]{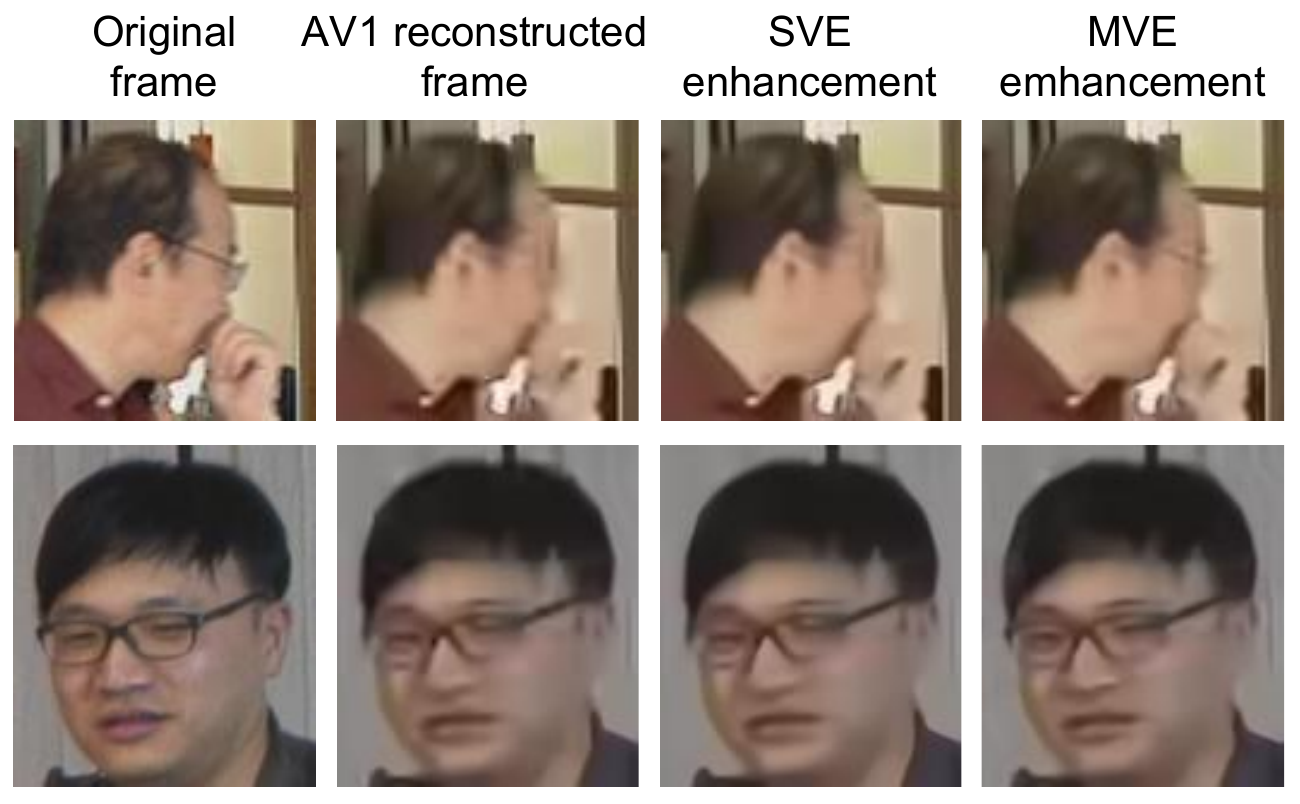}}
	\caption{\textbf{Qualitative Visualization} Zoomed-in snapshots of reconstructed frames for the AV1 baseline, SVE and MVE filtered restoration, as well as the ground truth label.}
	\label{fig:sec7_loop_filter_perceptual_comparison}
\end{figure}

\subsection{Discussion And Future Direction}
\label{subsec:discussion}
In this section, we proposed DNN-based approaches for video quality enhancement. For in-loop filtering, we developed a guided CNN framework to adapt pre-trained CNN models to various video contents. Under this framework, the guided CNN learns to project an input signal onto a subspace of dimension $\rm M$. The weighting parameters for a linear combination of these channels are explicitly signaled in the encoded bitstream to obtain the final restoration.
For post filtering, we devised a spatio-temporal multi-frame architecture to alleviate the compression artifacts. A two-step scheme is adopted in which optical flow is first obtained for accurate motion estimation/compensation, and then a wide activation residual network called WARN is designed for information fusion and quality enhancement. 
Our proposed enhancement approaches can be implemented on different CNN architectures.

The quality of enhanced frames plays a significant role for overall coding performance, since they serve as reference frames for the motion estimation of subsequent frames. Our future work will investigate the joint effect of in-loop filtering and motion estimation on reference frames to exploit the inherent correlations of these coding tools, which could further improve coding performance.

	\section{Discussion and Conclusion}
\label{sec:conclusion}
{As an old Chinese saying goes, ``A journey of a thousand miles begins with a single step.'' This is particularly true in the realm of technological advancement. Both the fields of video compression and machine learning have been established for many decades, but until recently, they evolved separately in both academic explorations and industrial practice.}

{Lately, however, we have begun to witness the interdisciplinary advancements yielded by the proactive application of deep learning technologies~\cite{lecun2015deep} into video compression systems. Benefits of these advances include remarkable improvements in performance in many technical aspects. To showcase the remarkable products of this disciplinary cross-pollination, we have identified three major functional blocks in a practical video system, {\it e.g.}, pre-processing, coding, post-processing. We then reviewed related studies and publications to help the audience familiarize themselves with these topics. Finally, we presented three case studies to highlight the state-of-the-art efficiency resulting from the application of DNNs to video compression systems, which demonstrates this avenue of exploration's great potential to bring about a new generation of video techniques, standards, and products.}

{Though this article presents separate DNN-based case studies for pre-processing, coding, and post-processing, we believe that a fully end-to-end DNN model could potentially offer a greater improvement in performance, while enabling more functionalities.  For example, Xia {\it et al.}~\cite{xia2020object} applied deep object segmentation in pre-processing, and used it to guide neural video coding, demonstrating noticeable visual improvements at very low bit rates. Meanwhile, Lee {\it et al.}~\cite{lee2019hybrid} and others observed similar effects, when a neural adaptive filter was successfully used to further enhance neural compressed images. }

{Nevertheless, a number of open problems requiring substantial further study have been discovered. These include:}

\begin{itemize}
    \item Model Generalization: It is vital for DNN models to be generalizable to a wide variety of video content, different artifacts, {\it etc.} Currently, most DNN-based video compression techniques utilize supervised learning, which often demands a significant amount of labelled image/video data for the full spectrum coverage of aforementioned application scenarios. Continuously developing a large-scale dataset, such as the ImageNet\footnote{\url{http://www.image-net.org/}} presents one possible solution to this problem. An alternative approach may use more advanced techniques to alleviate uncertainty related to a limited training sample for model generalization. These techniques include (but are not limited to) few-shot learning~\cite{wang2020generalizing} and self-supervised learning~\cite{lecun2015deep}.
    \item Complexity: Existing DNN-based methods are mainly criticized for their unbearable complexity in both computational and spatial dimensions. Compared to conventional video codec, which requires tens of Kilobytes on-chip memory, most DNN algorithms require several Megabytes or even Gigabytes of memory space. On the other hand, although inference may be very fast, training could take hours, days or even weeks for converged and reliable models~\cite{Lu_2019_CVPR}. All of these issues present serious barriers to the market adoption of DNN-based tools, particularly on energy-efficient mobile platforms. One promising solution is to design specialized hardware for the acceleration of DNN algorithms~\cite{hennessy2019new}. Currently, neural processing units (NPU) have attracted significant attention, and have been gradually deployed in heterogeneous platforms ({\it e.g.}, Qualcomm AI Engine in the Snapdragon chip series, Neural Processor in Apple silicons, {\it etc.}) This paints a promising picture of a future in which DNN algorithms can be deployed on NPU-equipped devices at a massive scale. 
    \item {QoE Metric: Video quality matters. A video QoE metric that is better correlated with the human visual system is highly desirable, not only for quality evaluation, but also for loss control in DNN-based video compression. There has been notable development in both subjective and objective video quality assessments, yielding several well-known metrics, such as SSIM~\cite{wang2003msssim}, just-noticeable-distortion (JND)~\cite{yuan2019visual}, and VMAF~\cite{vmaf}, some of which are actively adopted for the evaluation of video algorithms, application products, {\it etc.} On the other hand, existing DNN-based video coding approaches can adaptively optimize the efficiency of a pre-defined loss function, such as MSE, SSIM, adversarial loss~\cite{liu2018deep}, VGG feature based semantic loss, {\it etc.} However, none of these loss functions has shown clear advantages. A unified, differentiable, and HVS-driven metric is of great importance for the capacity of DNN-based video coding techniques to offer perceptually better QoE.}
\end{itemize}
    
The exponential growth of Internet traffic, a majority of which involves videos and images, has been the driving force for the development of video compression systems. The availability of a vast amount of images through the Internet, meanwhile, has been critical for the renaissance of the field of machine learning. In this work, we show that recent progress in deep learning can, in return, improve video compression. These mutual positive feedbacks suggest that significant progress could be achieved in both fields when they are investigated together. Therefore, the approaches presented in this work could be the stepping stones for improving the compression efficiency in Internet-scale video applications.

From a different perspective, most compressed videos will be ultimately consumed by human beings or interpreted by machines, for subsequent task decisions. This is a typical computer vision (CV) problem, {\it i.e.}, content understanding and decisions for consumption or task-oriented application ({\it e.g.}, detection, classification, {\it etc}.) Existing approaches have performed these tasks by first decoding the video, and then examining the tasks via learned or rule-based methods based on decoded pixels.
Such separate processing, {\it e.g.}, video decoding followed by CV tasks, is relied upon mainly because traditional pixel-prediction based differential video compression methods break the spatio-temporal features that could be potentially helpful for vision tasks. In contrast, recent DNN-based video compression algorithms rely on the feature extraction, activation, suppression, and aggregation for more compact representation. For these reasons, it is expected that the CV tasks can be fulfilled in the compressive domain without bit decoding and pixel reconstruction. Our earlier attempts have shown very encouraging  gain in the accuracy of classification and retrieval in compressive formats, without resorting to the traditional feature-based approaches using decoded pixels, which we report in ~\cite{shen2018codedvision,liu2019codedretrieval}. 
Using powerful DNNs to unify video compression and computer vision techniques is an exciting new field. It is also worth noting that the ISO/IEC MPEG is now actively working on a new project called ``Video Coding for Machine" (VCM)\footnote{\url{https://mpeg.chiariglione.org/standards/exploration/video-coding-machines}}, with emphasis on exploring video compression solutions for both human perception and machine intelligence.

	\bibliographystyle{IEEEtran}
	\bibliography{texture,E2E-NVC}
\end{document}